\documentclass[aps,pra,twocolumn,superscriptaddress,longbibliography]{revtex4-1}
\usepackage{graphicx}
\usepackage{latexsym}
\usepackage{amssymb}
\usepackage{amsmath}
\usepackage{amsfonts}
\usepackage{upgreek}
\usepackage{bm}
\usepackage{multirow}
\usepackage{color}
\usepackage[T1]{fontenc}
\usepackage{hyperref}
\hypersetup{
colorlinks = true,
linkcolor = [rgb]{0.70,0.13,0.13},
citecolor = [rgb]{0.13,0.55,0.13},
urlcolor  = [rgb]{0.25, 0.41, 0.88}}
\newcommand{\ket}[1]{|#1\rangle}

\newcommand{\braket}[2]{\langle#1|      #2\rangle}

\begin{document}

\title{Fidelity as a probe for a deconfined quantum critical point}
\author{Gaoyong Sun}
\thanks{Corresponding author: gysun@nuaa.edu.cn}
\affiliation{College of Science, Nanjing University of Aeronautics and Astronautics, Nanjing, 211106, China}
\author{Bo-Bo Wei}
\thanks{Corresponding author: weibobo@cuhk.edu.cn}
\affiliation{School of Science and Engineering, The Chinese University of Hong Kong, Shenzhen, Shenzhen 518172, China}
\affiliation{Center for Quantum Computing, Peng Cheng Laboratory, Shenzhen 518055, China}
\author{Su-Peng Kou}
\affiliation{Department of Physics, Beijing Normal University, Beijing 100875, China}

\begin{abstract}
Deconfined quantum critical point was proposed as a second-order quantum phase transition between two broken symmetry phases beyond the Landau-Ginzburg-Wilson paradigm.
However, numerical studies cannot completely rule out a weakly first-order transition because of strong violations of finite-size scaling.
We demonstrate that the fidelity is a simple probe to study deconfined quantum critical point.
We study the ground-state fidelity susceptibility close to the deconfined quantum critical point in a spin chain
using the large-scale finite-size density matrix renormalization group method.
We find that the finite-size scaling of the fidelity susceptibility obeys the conventional scaling behavior for continuous phase transitions, supporting the deconfined quantum phase transition is continuous.
We numerically determine the deconfined quantum critical point and the associated correlation length critical exponent from the finite-size scaling theory of the fidelity susceptibility.
Our results are consistent with recent results obtained directly from the matrix product states for infinite-size lattices using others observables.
Our work provides a useful probe to study critical behaviors at deconfined quantum critical point from the concept of quantum information.

\end{abstract}

\maketitle

\section{Introduction}
Many quantum phase transitions \cite{Sachdev1999} in strongly correlated many-body systems can be described by some order parameters according to the Landau-Ginzburg-Wilson (LGW) paradigm.
For continuous phase transitions, the behavior close to quantum critical point, such as the quantum critical point and the universal critical exponents, can be well described by the renormalization group theory \cite{Wilson1974,Wilson1975}. For a finite-size system, there is no phase transitions. But the critical point and the universal critical exponents of the phase transitions can be obtained from observables of finite size systems through the finite-size scaling theory \cite{Fisher1972,Fisher1974}.
In the LGW description, two spontaneous symmetry breaking phases would undergo either a first-order phase transition, or two phase transitions with an intermediate region between them.

The deconfined quantum critical point (DQCP)~\cite{Senthil2004,Senthil2004PRB} was proposed as an example with a direct second-order quantum phase transition between two broken symmetry phases, which is beyond the LGW paradigm.
A large number of two dimensional models were proposed to exhibit deconfined quantum phase transitions~\cite{Sandvik2007,Melko2008,Charrier2008,Kuklov2008,Chen2009,Lou2009,Charrier2010,Banerjee2010,Sandvik2010,Nahum2011,
Bartosch2013,Harada2013,Chen2013,Block2013,Sreejith2015,Nahum2015,Nahum2015PRL,Shao2016,Shao2017,Qin2017,Sato2017,Ma2018,Zhao2018,Serna2019,Ippoliti2018,Zhang2018}. However, the nature of the phase transition in these two-dimensional models is still under debate because of violations of finite-size scaling~\cite{Shao2016}, which was unexpected in deconfined quantum critical theory.
Two possibilities exist for the quantum phase transitions in the aforementioned two-dimensional models, they are either a weakly first-order phase transition described by the LGW paradigm or a second-order phase transition predicted by the deconfined quantum critical theory. To clarify the nature the deconfined quantum phase transitions~\cite{Shao2016}, it well deserves to investigate the behaviors of much more quantities beyond the traditional ones of correlation functions at the DQCP. In addition, the numerical simulations are very difficult to perform for two dimensional systems, especially for frustrated systems where the quantum Monto Carlo could fail due to sign problems \cite{Shao2016}.
While unfortunately most of designed models hosting DQCP are frustrated quantum magnets \cite{Shao2016}.
Recently, simple spin models~\cite{Patil2018,Jiang2019,Mudry2019,Roberts2019,Huang2019} are proposed to explore an analog of DQCP~\cite{Affleck1987,Sengupta2002,Sandvik2004},
which allows for easily numerical simulations with a high accuracy.

\begin{figure}
\includegraphics[width=8.6cm]{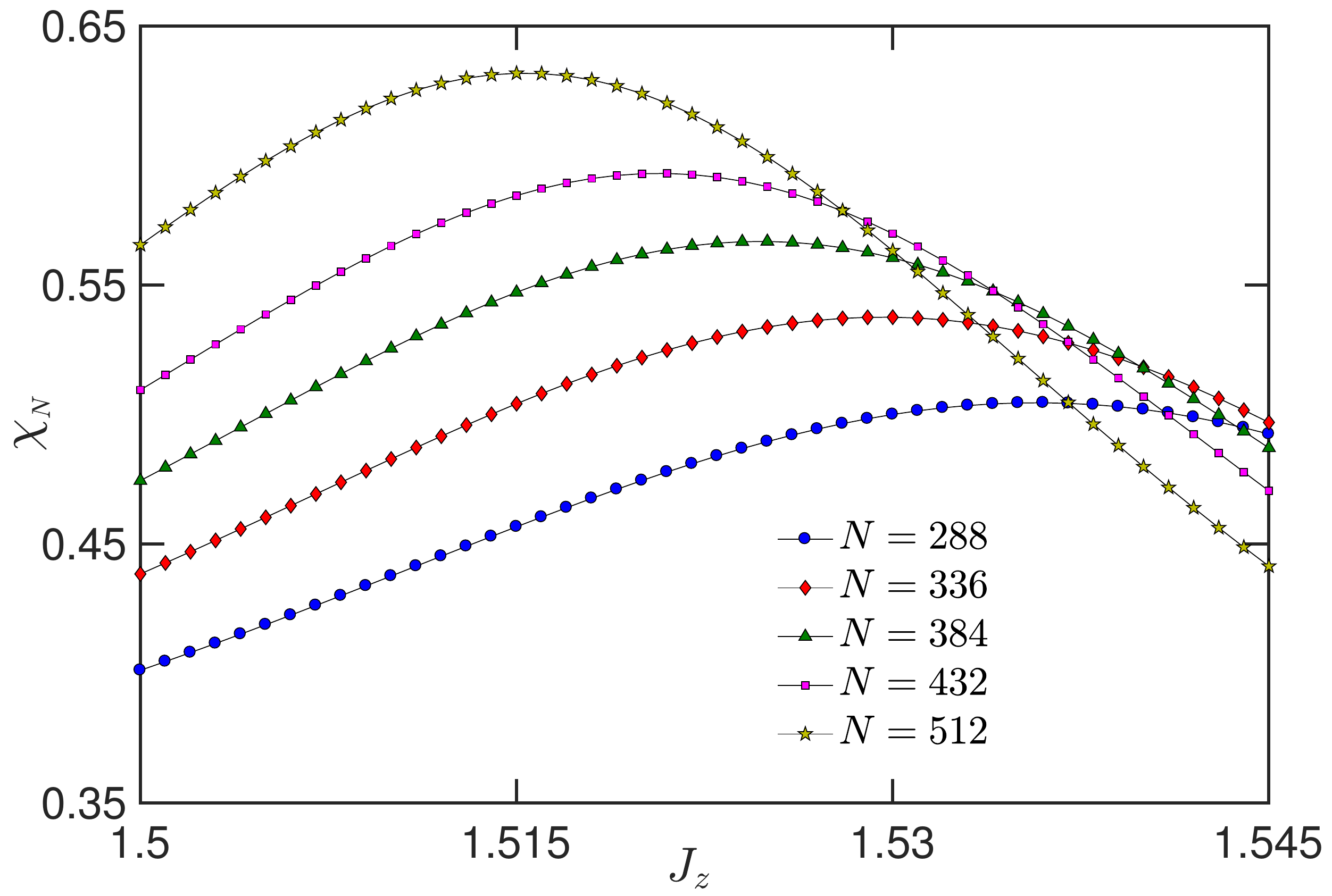}
\caption{(Color online)
Fidelity susceptibility per site $\chi_N$ near the DQCP of the spin chain as a function 
of $J_z$ for $N = 288$ (blue dot), $N=336$ (red diamond), $N=384$ (green triangle), $N=432$ (magenta square), $N=512$ (yellow star) lattice sites respectively. 
We fix $J_x=1$, $K_x=1/2$, $K_z=1/2$. Symbols denote DMRG numerical results and the solid lines are for eye guide.}
\label{FSfig}
\end{figure}

In this paper, we investigate the finite-size scaling of the fidelity susceptibility in a spin chain model which presents an analogy of DQCP using the finite-size density matrix renormalization group (DMRG) method \cite{White1992,Schollwock2005} based on the matrix product states \cite{Verstraete2004,Schollwock2011}.
We demonstrate that the fidelity is a simple probe to study the DQCP. Surprisingly, we find that the finite-size scaling of ground-state fidelity susceptibility obeys the conventional scaling behavior,
strongly supporting the phase transition is continuous. We extract the quantum critical point and the correlation length critical exponent of the deconfined quantum phase transitions using different scaling approaches.
The results we obtained agree with each other and are also consistent with recent results obtained directly from infinite-size systems.

\begin{figure}
\includegraphics[width=8.7cm]{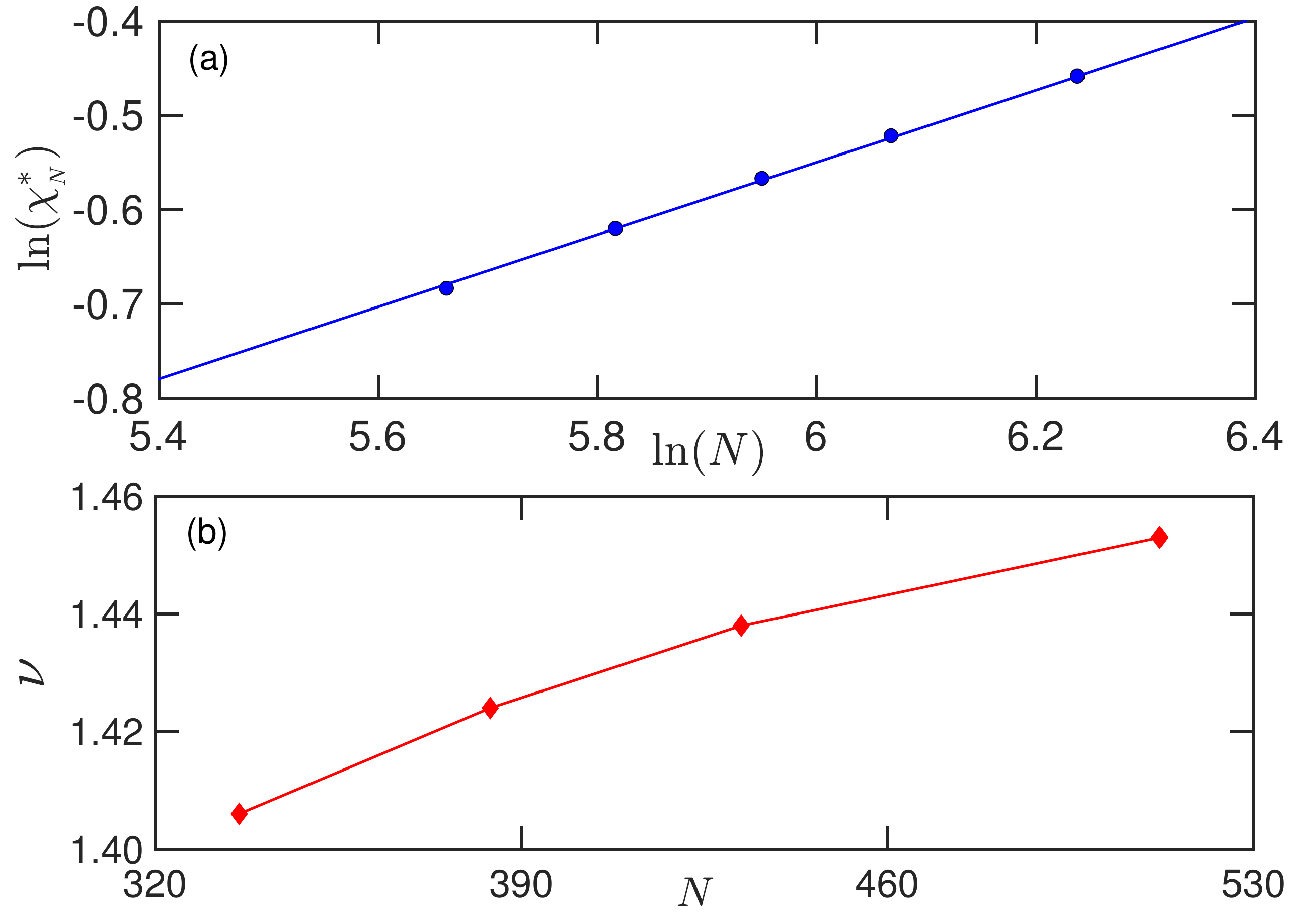}
\caption{(Color online)
Fidelity susceptibility peak and critical exponent $\nu$.
(a) The scaling of the amplitude of fidelity susceptibility per site $\chi_N^{\ast}$ at the peak position $J_z^{\ast}$ of the spin chain
as a function of the lattice size with $J_x=1$, $K_x=1/2$, $K_z=1/2$. Blue circle symbols denote DMRG numerical results and the solid line shows the linear fit from the data points.
The exponent of the correlation function $\nu=1.446$ is obtained from the linear fit. 
(b) The fitted critical exponent $\nu$ as a function of the maximum system size used in fitting. One can see a drift of critical exponents for bigger system sizes.}
\label{FSpeakfig}
\end{figure}

This paper is organized as follows.
In Sec. \ref{sec:model}, we briefly introduce the quantum spin chain model with DQCP.
In Sec. \ref{sec:FS}, we review the fidelity, fidelity susceptibility and their finite-size scaling behaviors.
In Sec. \ref{sec:DQCP}, we present the numerical results and discuss the finite-size scaling behaviors of the fidelity susceptibility near DQCP.
Finally, in Sec. \ref{sec:Con}, we give a discussion and a summary.

\section{Model}
\label{sec:model}
The model we considered here is a simple spin chain proposed recently~\cite{Jiang2019,Roberts2019,Huang2019} with the Hamiltonian
\begin{align}
H = \sum_{i} {}& (-J_{x} \sigma^{x}_{i} \sigma^{x}_{i+1} - J_{z} \sigma^{z}_{i} \sigma^{z}_{i+1} \nonumber \\
    {}& + K_{x} \sigma^{x}_{i} \sigma^{x}_{i+2}+K_{z} \sigma^{z}_{i} \sigma^{z}_{i+2}).
\label{Ham}
\end{align}
Here $\sigma^{\alpha}_{i}$ with $\alpha=x,z$ are the Pauli matrices at the $i$-th site along the $x$ and $z$ directions, and $J_{\alpha}$ and $K_{\alpha}$ are respectively the nearest-neighbor and next-nearest-neighbor spin-spin coupling constant. The model has the $\mathbb{Z}_2 \times \mathbb{Z}_2$ symmetry, the translation symmetry $\sigma^{\alpha}_{i} \rightarrow \sigma^{\alpha}_{i+1}$,
and the inversion symmetry $\sigma^{\alpha}_{i} \rightarrow \sigma^{\alpha}_{-i+1}$ \cite{Jiang2019}.
The critical values and critical exponents vary according to the couplings $J_{\alpha}$ and $K_{\alpha}$.
In the following, we simply choose $J_{x}=1,K_{x}=1/2,K_{z}=1/2$ and $J_{z}>0$ as studied in \cite{Huang2019}.
For $J_z=1$, the ground state is an exact valence bond solid (VBS) dimerized state (also called Majumdar-Ghosh state) \cite{Roberts2019}.
Increasing $J_z$, the ground state will become a ferromagnet phase (zFM).
The transition between VBS phase and zFM phase is a second-order quantum phase transition predicted by deconfined quantum critical theory.

In the following, we will study the finite-size scaling of the ground-state fidelity susceptibility near the critical point of this spin chain.
We support that the transition is a second-order continuous quantum phase transition by the finite-size scaling of the fidelity susceptibility.
Moreover, we argue that finite-size scaling of the fidelity susceptibility in the spin chain obeys a conventional scaling behavior,
which is the same as the finite-size scaling behaviors of other second-order transitions~\cite{Gu2010}.

\section{fidelity and fidelity susceptibility}
\label{sec:FS}
Given a general Hamiltonian $H(\lambda)=H_0 + \lambda H_1$ with $\lambda$ being a driving parameter, the ground-state fidelity is defined as the absolute value of the overlap between two ground-state wave functions~\cite{Zanardi2006},
\begin{align}
F(\lambda,\lambda+\delta \lambda)=|\braket{\psi_{0}(\lambda)}{\psi_{0}(\lambda +\delta \lambda)}|,
\end{align}
where $\ket{\psi_0(\lambda)}$ is the ground-state wave function of the Hamiltonian $H(\lambda)$,
and $\delta \lambda$ is a small change of parameter $\lambda$. Expanding the fidelity $F(\lambda,\lambda+\delta \lambda)$ up to second-order in small deviation $\delta\lambda$,
\begin{align}
F(\lambda,\lambda+\delta \lambda)=1-\frac{1}{2}\chi_F(\lambda) \delta \lambda^2,
\end{align}
we get the fidelity susceptibility $\chi_F(\lambda)$ as~\cite{You2007}
\begin{align}
\chi_F(\lambda) = \lim_{\delta \lambda \rightarrow 0} \frac{2(1-F(\lambda,\lambda+\delta \lambda))}{\delta \lambda^2}.
\end{align}
Since the overlap of two different ground states tends to zero, the fidelity susceptibility for finite systems will usually reach a maximum at a particular driving field $\lambda_{*}$ which is close to the critical point.
Therefore the fidelity susceptibility has been used to detect quantum phase transitions, including second-order phase transitions \cite{Zanardi2006,You2007,Venuti2007,Chen2008,Gu2008,Yang2008,Kwok2008,Gong2008,Yu2009,Schwandt2009,Gu2010,Albuquerque2010,Rams2011,Li2012,Victor2012,Rigol2013,Damski2013,Damski2014,Sun2017,Wei2018,Zhu2018,Lu2018,Wei2019}
and topological Berezinsky-Kosterlitz-Thouless (BKT) transitions \cite{Yang2007,Fjestad2018,Langari2012,Carrasquilla2013,Lacki2014,Sun2015}.
We note that the peak of fidelity susceptibility around BKT transitions may shift from the quantum critical point due to a non-trivial subleading term \cite{Cincio2019}.

\begin{figure}
\includegraphics[width=8.6cm]{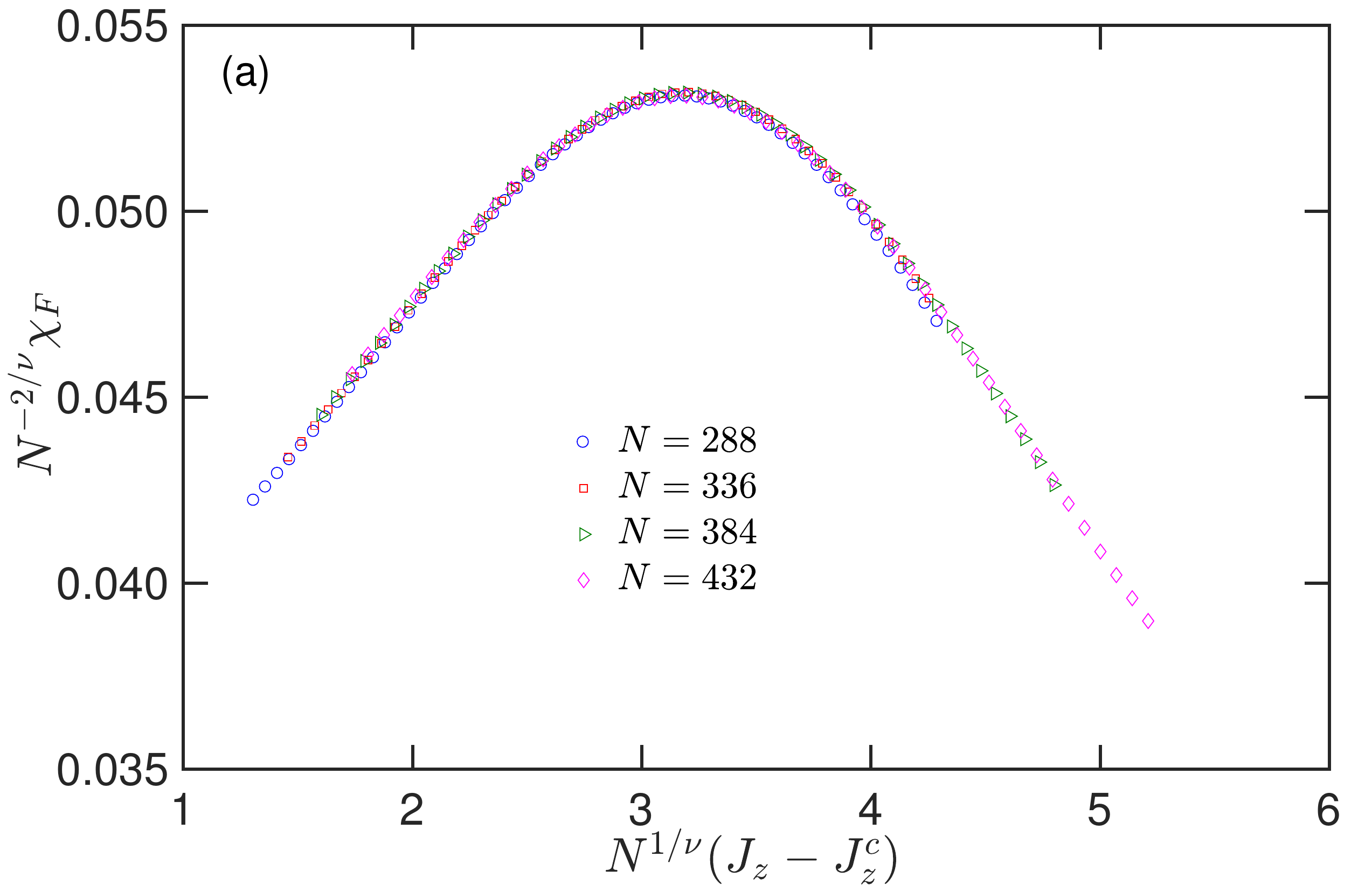}
\includegraphics[width=8.6cm]{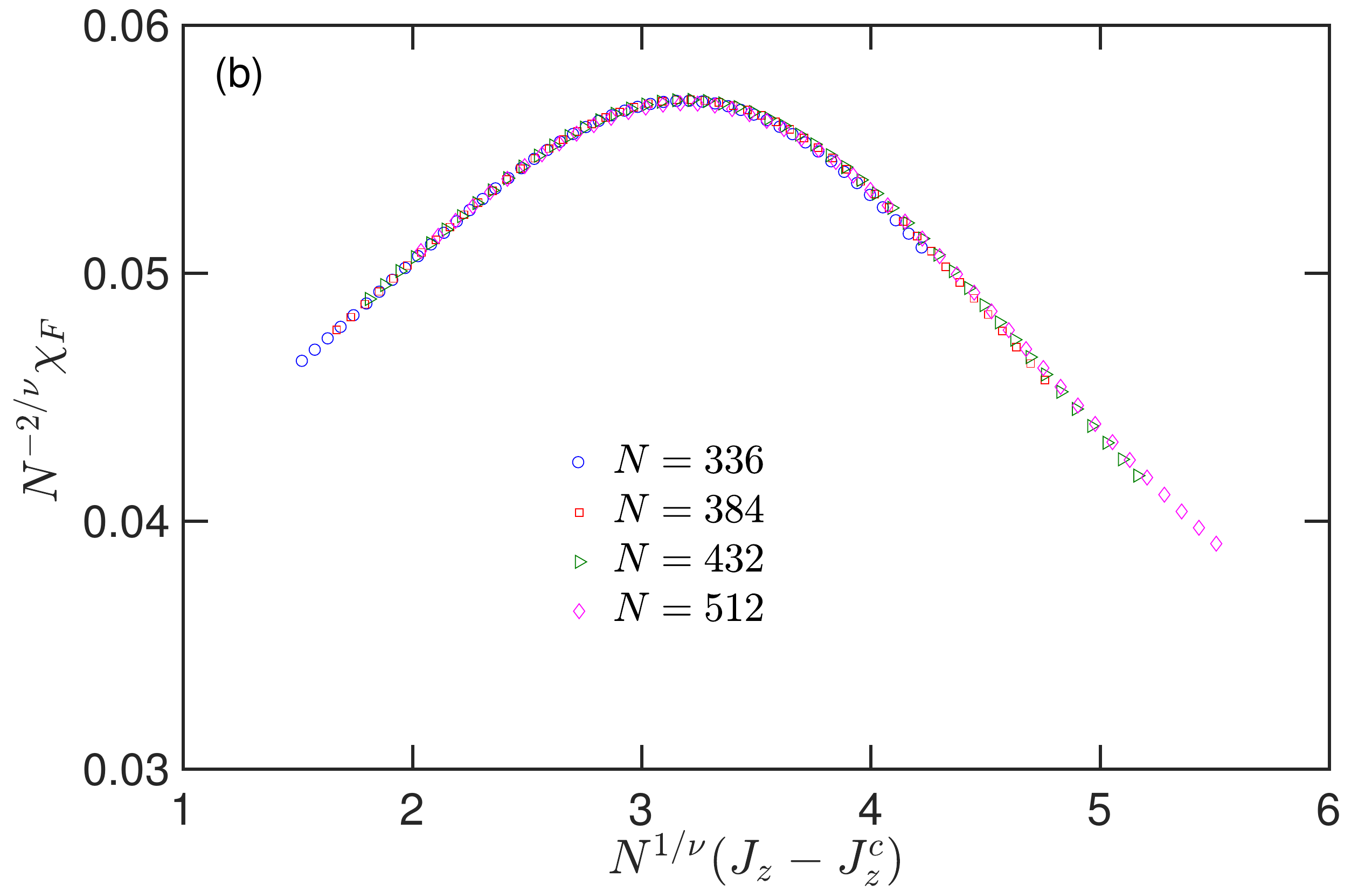}
\caption{(Color online)  Data collapse of the fidelity susceptibility $\chi_F=N \cdot \chi_L$ shown in Fig.\ref{FSfig}.
(a). Data collapse for $N = 288,336,384,432$ sites, where the exponent of the correlation function $ \nu = 1.431$ and the critical value $J_z^{c} = 1.475$ are chosen to achieve perfect data collapse.
(b). Data collapse for $N = 336,384,432, 512$ sites, where the critical exponent $ \nu = 1.443$ and the critical value $J_z^{c} = 1.473$ are chosen to achieve perfect data collapse. Symbols denote the rescaled DMRG results.}
\label{Collapsefig}
\end{figure}

For a continuous second-order transition, it was shown that the fidelity susceptibility $\chi_F$ near the critical point scales with the system size $N$ as~\cite{Gu2010,Albuquerque2010}
\begin{align}
\chi_F \propto N^{2 / \nu},
\label{FSscale}
\end{align}
with $\nu$ being the correlation length critical exponent. We can extract the exponent $\nu$ by fitting the scaling function from Eq.(\ref{FSscale}).
For instance, there is an exact solution of fidelity susceptibility, $\chi_F=\frac{N^2}{32} - \frac{N}{32}$, at critical point for one dimensional quantum Ising model~\cite{Damski2013}.
If $N \gg 1$, one can ignore the subleading term $- \frac{N}{32}$, the fidelity susceptibility will scale as $\chi_F \propto N^2$,
implying that the exponent $\nu=1$ for the quantum Ising chain from Eq.(\ref{FSscale}).
We note that: (1) the scaling described by Eq.(\ref{FSscale}) is only correct in the vicinity of critical value $\lambda_c$.
For any finite systems, the critical value $\lambda_c$ means the peak position $\lambda_{\ast}$ corresponding to the maximum of the fidelity susceptibility;
(2) usually there is an unknown subleading term for the fidelity susceptibility, i.e. the $- \frac{N}{32}$ term for the quantum Ising chain.
Both facts will demand numerical simulations for very large system sizes. Therefore there will be a small drift when extracting the exponent $\nu$ using different lattice sizes.
However we note that the drift coming from fidelity susceptibility should be different from the drift found for DQCP \cite{Nahum2015,Zhang2018} due to the anomalous finite-size scaling of physical quantities.
Because such a drift comes from the subleading term of the fidelity susceptibility and in principle can occur for all second-order transitions
and can be ignored when fitting the data up to hundreds or thousands of lattice sizes.
While for the DQCP, the drifts due to the anomalous scaling behaviors is argued to be the finite-size effects of dangerously irrelevant operators \cite{Nahum2015,Zhang2018}.

\section{finite-size scaling near DQCP}
\label{sec:DQCP}
In the following, we perform the DMRG simulations based on the matrix product states and we use the open boundary conditions to get a better accuracy.
We compute the fidelity susceptibility $\chi_F$ for $N=288,336,384,432, 512$ sizes and keep $M=300$ to $M=500$ states
with the step of driving parameters $dJ_z=10^{-3}$ and $dJ_z=10^{-4}$ during the simulations.
We find that the ground state energy and the fidelity susceptibility converge respectively up to the order $10^{-8}$ and $10^{-5}$.
Fig.\ref{FSfig} shows the fidelity susceptibility per site $\chi_N = \chi_F/N$ as a function of control parameter $J_z$ for different lattice sizes. One can clearly see that there is a peak in the fidelity susceptibility located in the zFM phase and the peak of fidelity susceptibility per size $\chi_N$ increases and moves towards to the VBS phase with the increase of the system sizes $N$.
We determine the correlation length critical exponent $\nu=1.446$ using the maximum of the fidelity susceptibility with $L=336,384,432,512$ as shown in Fig.\ref{FSpeakfig}(a) according to Eq.(\ref{FSscale}).
In Fig.\ref{FSpeakfig}(b), we show the fitted critical exponents as a function of the largest system sizes used in fitting. We get one critical exponent $\nu$ by data in three consecutive system sizes, 
such as the first data point in Fig.\ref{FSpeakfig}(b) is obtained by fitting the maximum of fidelity susceptibility of three system sizes $N = 240, 288, 336$ respectively, 
and the last data point in Fig.\ref{FSpeakfig}(b) is obtained by fitting the maximum of fidelity susceptibility of three system sizes $N = 384, 432, 512$ respectively. 
One can see that a drift in the fitted critical exponent with a difference up to $2 \%$ for the smallest and biggest sizes is obtained. 
We note that these small drifts come from the subleading term of fidelity susceptibility, which are fundamentally different from the drifts due to anomalous scaling behaviors of two-dimensional DQCPs as we mentioned above.

Alternatively, the correlation length critical exponent $\nu$ can be determined by finite-size scaling of $\chi_{F}$ \cite{Albuquerque2010,Sun2017,Zhu2018,Wei2019} at continuous phase transitions,
\begin{align}
\chi_{F}(h) = L^{(2/ \nu)}f_{\chi_{F}}(L^{1/\nu}|J_z-J_z^{c}|)
\label{eqFScollapse}
\end{align}
Eq.~\eqref{eqFScollapse} tells us that if we plot $\chi_F/L^{2/\nu}$ as a function of scaled parameter $L^{1/\nu}|J_z-J_z^{c}|$ for different system sizes, all curves collapse into a single one if $\nu$ is properly chosen. As shown in Fig.\ref{Collapsefig}, we plot the scaled fidelity susceptibility $L^{-2/ \nu}\chi_F$ as a function of $L^{1/\nu}|J_z-J_z^{c}|$.
We adjust the parameters $\nu$ and $J_z^{c}$ until all the curves collapse perfectly. The critical exponent $\nu = 1.431$ and critical point $J_z^{c} = 1.475$ are obtained from $L=288,336,384,432$ as shown in Fig.\ref{Collapsefig}(a). To study the drifts of critical exponent and critical point due to finite-size scaling, we perform the other data collapse using $L=336,384,432, 512$,
and obtain the critical exponent $\nu = 1.443 \pm 0.009$ and critical point $J_z^{c} = 1.473 \pm 0.003$.
The critical exponent $\nu$ extracted from the above two independent methods agree with each other very well, indicating our results are trustable.
In addition, the critical exponent $\nu$ and the critical point $J_z^c$ are consistent with results obtained directly from the matrix product states for infinite systems in Ref.\cite{Huang2019}.

It is known that a weakly first-order transition with a huge but finite correlation length $\xi$ can also show a "pseudoscaling" with a nice data collapse \cite{Wang2017PRX,Iino2019}.
In order to rule out the weakly first-order transition, one has to compute the correlation length $\xi$.
It is shown in Ref.\cite{Roberts2019,Huang2019} that the correlation length $\xi$ diverges. Hence, such a conventional finite-size scaling behavior of fidelity susceptibility indicates that the phase transition at DQCP is second-order. 
It is an open question that why there are the anomalous scaling behavior at two-dimensional DQCPs.
Here, we found that the fidelity susceptibility at one-dimensional DQCP obeys the usual scaling behaviors. We note that the conventional finite-size scaling may only exist in one dimensional systems.
For two-dimensional DQCP, based on the J-Q model \cite{Sandvik2010} for spins $S = 1/2$, it was found that there exist two divergent length scales, that may correspond to the length and the width of the strings for spinon excitations. See Ref.\cite{Shao2016}. As a result, it is difficult to learn a two-dimensional DQCP with the help of a variable, for example fidelity susceptibility to characterize the two divergent length scales. 
In addition, from point view of numerical simulations, although the DMRG approach is very successful to study the one-dimensional DQCP, it is still of challenge to apply it to study two-dimensional DQCP.
We provide an example to understand the one-dimensional DQCPs from aspect of the quantum information.
It would be interesting to investigate whether other geometric tensors, such as quantum Fisher information~\cite{Zoller2016}, the geometric phase~\cite{Carollo2005} and Loschmidt echoes~\cite{Hwang2019} obey the conventional scaling behaviors at one-dimensional DQCPs.

\section {Conclusion}
\label{sec:Con}
In summary, we have shown that the fidelity susceptibility can be used as a probe for detecting the DQCP.
We have extracted the critical point and the correlation length critical exponent of the deconfined quantum phase transitions from the finite-size scaling of the fidelity susceptibility.
More importantly, we have shown that the fidelity susceptibility obeys the conventional finite-size scaling behaviors at the DQCP, which supports that the DQCP is of second order phase transitions. 
It is interesting to investigate wether the deconfined critical theory can prove that the fidelity susceptibility indeed obey the conventional scaling behaviors,
and whether other geometric tensors, such as the quantum Fisher information~\cite{Zoller2016}, the geometric phase~\cite{Carollo2005} and the Loschmidt echoes~\cite{Hwang2019} would obey the conventional finite-size scaling theory. Meanwhile, it would be very important to study the finite-size scaling of fidelity susceptibility for two dimensional systems with DQCP using the quantum Monte Carlo method \cite{Wang2015,Cai2019}.
In addition, it was shown that~\cite{Gu2014,You2015} the fidelity susceptibility is connected to dynamical structure factor which can be measured experimentally in the linear response regime, thus
experimental measurement of the fidelity susceptibility for deconfined quantum phase transitions may be performed in near future.

{\it Note added}.- After the submission of our paper, we became aware of a work on conventional the finite-size scaling of the one-dimensional DQCP in the same model using order parameters \cite{Luo2019}.

\begin{acknowledgments}
G. S. is appreciative of support from the NSFC under the Grant No. 11704186 and the startup Fund of Nanjing University of Aeronautics and Astronautics under the Grant No. YAH17053.
B. B. W. is appreciative of support from the NSFC under the Grant No. 11604220 and the President's Fund of The Chinese University of Hong Kong, Shenzhen.
S. P. K. is appreciative of support from the NSFC under the Grant No. 11674026.
Numerical simulations were performed on the clusters at National Supercomputing Center in Shenzhen and Nanjing University of Aeronautics and Astronautics.
\end{acknowledgments}

\bibliographystyle{apsrev4-1}
\bibliography{ref}

\begin{thebibliography}{85}%
\makeatletter
\providecommand \@ifxundefined [1]{%
 \@ifx{#1\undefined}
}%
\providecommand \@ifnum [1]{%
 \ifnum #1\expandafter \@firstoftwo
 \else \expandafter \@secondoftwo
 \fi
}%
\providecommand \@ifx [1]{%
 \ifx #1\expandafter \@firstoftwo
 \else \expandafter \@secondoftwo
 \fi
}%
\providecommand \natexlab [1]{#1}%
\providecommand \enquote  [1]{``#1''}%
\providecommand \bibnamefont  [1]{#1}%
\providecommand \bibfnamefont [1]{#1}%
\providecommand \citenamefont [1]{#1}%
\providecommand \href@noop [0]{\@secondoftwo}%
\providecommand \href [0]{\begingroup \@sanitize@url \@href}%
\providecommand \@href[1]{\@@startlink{#1}\@@href}%
\providecommand \@@href[1]{\endgroup#1\@@endlink}%
\providecommand \@sanitize@url [0]{\catcode `\\12\catcode `\$12\catcode
  `\&12\catcode `\#12\catcode `\^12\catcode `\_12\catcode `\%12\relax}%
\providecommand \@@startlink[1]{}%
\providecommand \@@endlink[0]{}%
\providecommand \url  [0]{\begingroup\@sanitize@url \@url }%
\providecommand \@url [1]{\endgroup\@href {#1}{\urlprefix }}%
\providecommand \urlprefix  [0]{URL }%
\providecommand \Eprint [0]{\href }%
\providecommand \doibase [0]{http://dx.doi.org/}%
\providecommand \selectlanguage [0]{\@gobble}%
\providecommand \bibinfo  [0]{\@secondoftwo}%
\providecommand \bibfield  [0]{\@secondoftwo}%
\providecommand \translation [1]{[#1]}%
\providecommand \BibitemOpen [0]{}%
\providecommand \bibitemStop [0]{}%
\providecommand \bibitemNoStop [0]{.\EOS\space}%
\providecommand \EOS [0]{\spacefactor3000\relax}%
\providecommand \BibitemShut  [1]{\csname bibitem#1\endcsname}%
\let\auto@bib@innerbib\@empty
\bibitem [{\citenamefont {Sachdev}(1999)}]{Sachdev1999}%
  \BibitemOpen
  \bibfield  {author} {\bibinfo {author} {\bibfnamefont {S.}~\bibnamefont
  {Sachdev}},\ }\href@noop {} {\emph {\bibinfo {title} {Quantum phase
  transitions}}}\ (\bibinfo  {publisher} {Cambridge University Press,
  Cambridge},\ \bibinfo {year} {1999})\BibitemShut {NoStop}%
\bibitem [{\citenamefont {Wilson}\ and\ \citenamefont
  {Kogut}(1974)}]{Wilson1974}%
  \BibitemOpen
  \bibfield  {author} {\bibinfo {author} {\bibfnamefont {K.~G.}\ \bibnamefont
  {Wilson}}\ and\ \bibinfo {author} {\bibfnamefont {J.}~\bibnamefont {Kogut}},\
  }\href {\doibase 10.1016/0370-1573(74)90023-4} {\bibfield  {journal}
  {\bibinfo  {journal} {Physics reports}\ }\textbf {\bibinfo {volume} {12}},\
  \bibinfo {pages} {75} (\bibinfo {year} {1974})}\BibitemShut {NoStop}%
\bibitem [{\citenamefont {Wilson}(1975)}]{Wilson1975}%
  \BibitemOpen
  \bibfield  {author} {\bibinfo {author} {\bibfnamefont {K.~G.}\ \bibnamefont
  {Wilson}},\ }\href {\doibase 10.1103/RevModPhys.47.773} {\bibfield  {journal}
  {\bibinfo  {journal} {Reviews of modern physics}\ }\textbf {\bibinfo {volume}
  {47}},\ \bibinfo {pages} {773} (\bibinfo {year} {1975})}\BibitemShut
  {NoStop}%
\bibitem [{\citenamefont {Fisher}\ and\ \citenamefont
  {Barber}(1972)}]{Fisher1972}%
  \BibitemOpen
  \bibfield  {author} {\bibinfo {author} {\bibfnamefont {M.~E.}\ \bibnamefont
  {Fisher}}\ and\ \bibinfo {author} {\bibfnamefont {M.~N.}\ \bibnamefont
  {Barber}},\ }\href {\doibase 10.1103/PhysRevLett.28.1516} {\bibfield
  {journal} {\bibinfo  {journal} {Physical Review Letters}\ }\textbf {\bibinfo
  {volume} {28}},\ \bibinfo {pages} {1516} (\bibinfo {year}
  {1972})}\BibitemShut {NoStop}%
\bibitem [{\citenamefont {Fisher}(1974)}]{Fisher1974}%
  \BibitemOpen
  \bibfield  {author} {\bibinfo {author} {\bibfnamefont {M.~E.}\ \bibnamefont
  {Fisher}},\ }\href {\doibase 10.1103/RevModPhys.46.597} {\bibfield  {journal}
  {\bibinfo  {journal} {Reviews of Modern Physics}\ }\textbf {\bibinfo {volume}
  {46}},\ \bibinfo {pages} {597} (\bibinfo {year} {1974})}\BibitemShut
  {NoStop}%
\bibitem [{\citenamefont {Senthil}\ \emph
  {et~al.}(2004{\natexlab{a}})\citenamefont {Senthil}, \citenamefont
  {Vishwanath}, \citenamefont {Balents}, \citenamefont {Sachdev},\ and\
  \citenamefont {Fisher}}]{Senthil2004}%
  \BibitemOpen
  \bibfield  {author} {\bibinfo {author} {\bibfnamefont {T.}~\bibnamefont
  {Senthil}}, \bibinfo {author} {\bibfnamefont {A.}~\bibnamefont {Vishwanath}},
  \bibinfo {author} {\bibfnamefont {L.}~\bibnamefont {Balents}}, \bibinfo
  {author} {\bibfnamefont {S.}~\bibnamefont {Sachdev}}, \ and\ \bibinfo
  {author} {\bibfnamefont {M.~P.}\ \bibnamefont {Fisher}},\ }\href {\doibase
  10.1126/science.1091806} {\bibfield  {journal} {\bibinfo  {journal}
  {Science}\ }\textbf {\bibinfo {volume} {303}},\ \bibinfo {pages} {1490}
  (\bibinfo {year} {2004}{\natexlab{a}})}\BibitemShut {NoStop}%
\bibitem [{\citenamefont {Senthil}\ \emph
  {et~al.}(2004{\natexlab{b}})\citenamefont {Senthil}, \citenamefont {Balents},
  \citenamefont {Sachdev}, \citenamefont {Vishwanath},\ and\ \citenamefont
  {Fisher}}]{Senthil2004PRB}%
  \BibitemOpen
  \bibfield  {author} {\bibinfo {author} {\bibfnamefont {T.}~\bibnamefont
  {Senthil}}, \bibinfo {author} {\bibfnamefont {L.}~\bibnamefont {Balents}},
  \bibinfo {author} {\bibfnamefont {S.}~\bibnamefont {Sachdev}}, \bibinfo
  {author} {\bibfnamefont {A.}~\bibnamefont {Vishwanath}}, \ and\ \bibinfo
  {author} {\bibfnamefont {M.~P.}\ \bibnamefont {Fisher}},\ }\href {\doibase
  10.1103/PhysRevB.70.144407} {\bibfield  {journal} {\bibinfo  {journal}
  {Physical Review B}\ }\textbf {\bibinfo {volume} {70}},\ \bibinfo {pages}
  {144407} (\bibinfo {year} {2004}{\natexlab{b}})}\BibitemShut {NoStop}%
\bibitem [{\citenamefont {Sandvik}(2007)}]{Sandvik2007}%
  \BibitemOpen
  \bibfield  {author} {\bibinfo {author} {\bibfnamefont {A.~W.}\ \bibnamefont
  {Sandvik}},\ }\href {\doibase 10.1103/PhysRevLett.98.227202} {\bibfield
  {journal} {\bibinfo  {journal} {Physical review letters}\ }\textbf {\bibinfo
  {volume} {98}},\ \bibinfo {pages} {227202} (\bibinfo {year}
  {2007})}\BibitemShut {NoStop}%
\bibitem [{\citenamefont {Melko}\ and\ \citenamefont {Kaul}(2008)}]{Melko2008}%
  \BibitemOpen
  \bibfield  {author} {\bibinfo {author} {\bibfnamefont {R.~G.}\ \bibnamefont
  {Melko}}\ and\ \bibinfo {author} {\bibfnamefont {R.~K.}\ \bibnamefont
  {Kaul}},\ }\href {\doibase 10.1103/PhysRevLett.100.017203} {\bibfield
  {journal} {\bibinfo  {journal} {Physical review letters}\ }\textbf {\bibinfo
  {volume} {100}},\ \bibinfo {pages} {017203} (\bibinfo {year}
  {2008})}\BibitemShut {NoStop}%
\bibitem [{\citenamefont {Charrier}\ \emph {et~al.}(2008)\citenamefont
  {Charrier}, \citenamefont {Alet},\ and\ \citenamefont
  {Pujol}}]{Charrier2008}%
  \BibitemOpen
  \bibfield  {author} {\bibinfo {author} {\bibfnamefont {D.}~\bibnamefont
  {Charrier}}, \bibinfo {author} {\bibfnamefont {F.}~\bibnamefont {Alet}}, \
  and\ \bibinfo {author} {\bibfnamefont {P.}~\bibnamefont {Pujol}},\ }\href
  {\doibase 10.1103/PhysRevLett.101.167205} {\bibfield  {journal} {\bibinfo
  {journal} {Physical review letters}\ }\textbf {\bibinfo {volume} {101}},\
  \bibinfo {pages} {167205} (\bibinfo {year} {2008})}\BibitemShut {NoStop}%
\bibitem [{\citenamefont {Kuklov}\ \emph {et~al.}(2008)\citenamefont {Kuklov},
  \citenamefont {Matsumoto}, \citenamefont {Prokof'Ev}, \citenamefont
  {Svistunov},\ and\ \citenamefont {Troyer}}]{Kuklov2008}%
  \BibitemOpen
  \bibfield  {author} {\bibinfo {author} {\bibfnamefont {A.}~\bibnamefont
  {Kuklov}}, \bibinfo {author} {\bibfnamefont {M.}~\bibnamefont {Matsumoto}},
  \bibinfo {author} {\bibfnamefont {N.}~\bibnamefont {Prokof'Ev}}, \bibinfo
  {author} {\bibfnamefont {B.}~\bibnamefont {Svistunov}}, \ and\ \bibinfo
  {author} {\bibfnamefont {M.}~\bibnamefont {Troyer}},\ }\href {\doibase
  10.1103/PhysRevLett.101.050405} {\bibfield  {journal} {\bibinfo  {journal}
  {Physical review letters}\ }\textbf {\bibinfo {volume} {101}},\ \bibinfo
  {pages} {050405} (\bibinfo {year} {2008})}\BibitemShut {NoStop}%
\bibitem [{\citenamefont {Chen}\ \emph {et~al.}(2009)\citenamefont {Chen},
  \citenamefont {Gukelberger}, \citenamefont {Trebst}, \citenamefont {Alet},\
  and\ \citenamefont {Balents}}]{Chen2009}%
  \BibitemOpen
  \bibfield  {author} {\bibinfo {author} {\bibfnamefont {G.}~\bibnamefont
  {Chen}}, \bibinfo {author} {\bibfnamefont {J.}~\bibnamefont {Gukelberger}},
  \bibinfo {author} {\bibfnamefont {S.}~\bibnamefont {Trebst}}, \bibinfo
  {author} {\bibfnamefont {F.}~\bibnamefont {Alet}}, \ and\ \bibinfo {author}
  {\bibfnamefont {L.}~\bibnamefont {Balents}},\ }\href {\doibase
  10.1103/PhysRevB.80.045112} {\bibfield  {journal} {\bibinfo  {journal}
  {Physical Review B}\ }\textbf {\bibinfo {volume} {80}},\ \bibinfo {pages}
  {045112} (\bibinfo {year} {2009})}\BibitemShut {NoStop}%
\bibitem [{\citenamefont {Lou}\ \emph {et~al.}(2009)\citenamefont {Lou},
  \citenamefont {Sandvik},\ and\ \citenamefont {Kawashima}}]{Lou2009}%
  \BibitemOpen
  \bibfield  {author} {\bibinfo {author} {\bibfnamefont {J.}~\bibnamefont
  {Lou}}, \bibinfo {author} {\bibfnamefont {A.~W.}\ \bibnamefont {Sandvik}}, \
  and\ \bibinfo {author} {\bibfnamefont {N.}~\bibnamefont {Kawashima}},\ }\href
  {\doibase 10.1103/PhysRevB.80.180414} {\bibfield  {journal} {\bibinfo
  {journal} {Physical Review B}\ }\textbf {\bibinfo {volume} {80}},\ \bibinfo
  {pages} {180414} (\bibinfo {year} {2009})}\BibitemShut {NoStop}%
\bibitem [{\citenamefont {Charrier}\ and\ \citenamefont
  {Alet}(2010)}]{Charrier2010}%
  \BibitemOpen
  \bibfield  {author} {\bibinfo {author} {\bibfnamefont {D.}~\bibnamefont
  {Charrier}}\ and\ \bibinfo {author} {\bibfnamefont {F.}~\bibnamefont
  {Alet}},\ }\href {\doibase 10.1103/PhysRevB.82.014429} {\bibfield  {journal}
  {\bibinfo  {journal} {Physical Review B}\ }\textbf {\bibinfo {volume} {82}},\
  \bibinfo {pages} {014429} (\bibinfo {year} {2010})}\BibitemShut {NoStop}%
\bibitem [{\citenamefont {Banerjee}\ \emph {et~al.}(2010)\citenamefont
  {Banerjee}, \citenamefont {Damle},\ and\ \citenamefont
  {Alet}}]{Banerjee2010}%
  \BibitemOpen
  \bibfield  {author} {\bibinfo {author} {\bibfnamefont {A.}~\bibnamefont
  {Banerjee}}, \bibinfo {author} {\bibfnamefont {K.}~\bibnamefont {Damle}}, \
  and\ \bibinfo {author} {\bibfnamefont {F.}~\bibnamefont {Alet}},\ }\href
  {\doibase 10.1103/PhysRevB.82.155139} {\bibfield  {journal} {\bibinfo
  {journal} {Physical Review B}\ }\textbf {\bibinfo {volume} {82}},\ \bibinfo
  {pages} {155139} (\bibinfo {year} {2010})}\BibitemShut {NoStop}%
\bibitem [{\citenamefont {Sandvik}(2010)}]{Sandvik2010}%
  \BibitemOpen
  \bibfield  {author} {\bibinfo {author} {\bibfnamefont {A.~W.}\ \bibnamefont
  {Sandvik}},\ }\href {\doibase 10.1103/PhysRevLett.104.177201} {\bibfield
  {journal} {\bibinfo  {journal} {Physical review letters}\ }\textbf {\bibinfo
  {volume} {104}},\ \bibinfo {pages} {177201} (\bibinfo {year}
  {2010})}\BibitemShut {NoStop}%
\bibitem [{\citenamefont {Nahum}\ \emph {et~al.}(2011)\citenamefont {Nahum},
  \citenamefont {Chalker}, \citenamefont {Serna}, \citenamefont {Ortuno},\ and\
  \citenamefont {Somoza}}]{Nahum2011}%
  \BibitemOpen
  \bibfield  {author} {\bibinfo {author} {\bibfnamefont {A.}~\bibnamefont
  {Nahum}}, \bibinfo {author} {\bibfnamefont {J.}~\bibnamefont {Chalker}},
  \bibinfo {author} {\bibfnamefont {P.}~\bibnamefont {Serna}}, \bibinfo
  {author} {\bibfnamefont {M.}~\bibnamefont {Ortuno}}, \ and\ \bibinfo {author}
  {\bibfnamefont {A.}~\bibnamefont {Somoza}},\ }\href {\doibase
  10.1103/PhysRevLett.107.110601} {\bibfield  {journal} {\bibinfo  {journal}
  {Physical review letters}\ }\textbf {\bibinfo {volume} {107}},\ \bibinfo
  {pages} {110601} (\bibinfo {year} {2011})}\BibitemShut {NoStop}%
\bibitem [{\citenamefont {Bartosch}(2013)}]{Bartosch2013}%
  \BibitemOpen
  \bibfield  {author} {\bibinfo {author} {\bibfnamefont {L.}~\bibnamefont
  {Bartosch}},\ }\href {\doibase 10.1103/PhysRevB.88.195140} {\bibfield
  {journal} {\bibinfo  {journal} {Physical Review B}\ }\textbf {\bibinfo
  {volume} {88}},\ \bibinfo {pages} {195140} (\bibinfo {year}
  {2013})}\BibitemShut {NoStop}%
\bibitem [{\citenamefont {Harada}\ \emph {et~al.}(2013)\citenamefont {Harada},
  \citenamefont {Suzuki}, \citenamefont {Okubo}, \citenamefont {Matsuo},
  \citenamefont {Lou}, \citenamefont {Watanabe}, \citenamefont {Todo},\ and\
  \citenamefont {Kawashima}}]{Harada2013}%
  \BibitemOpen
  \bibfield  {author} {\bibinfo {author} {\bibfnamefont {K.}~\bibnamefont
  {Harada}}, \bibinfo {author} {\bibfnamefont {T.}~\bibnamefont {Suzuki}},
  \bibinfo {author} {\bibfnamefont {T.}~\bibnamefont {Okubo}}, \bibinfo
  {author} {\bibfnamefont {H.}~\bibnamefont {Matsuo}}, \bibinfo {author}
  {\bibfnamefont {J.}~\bibnamefont {Lou}}, \bibinfo {author} {\bibfnamefont
  {H.}~\bibnamefont {Watanabe}}, \bibinfo {author} {\bibfnamefont
  {S.}~\bibnamefont {Todo}}, \ and\ \bibinfo {author} {\bibfnamefont
  {N.}~\bibnamefont {Kawashima}},\ }\href {\doibase 10.1103/PhysRevB.88.220408}
  {\bibfield  {journal} {\bibinfo  {journal} {Physical Review B}\ }\textbf
  {\bibinfo {volume} {88}},\ \bibinfo {pages} {220408} (\bibinfo {year}
  {2013})}\BibitemShut {NoStop}%
\bibitem [{\citenamefont {Chen}\ \emph {et~al.}(2013)\citenamefont {Chen},
  \citenamefont {Huang}, \citenamefont {Deng}, \citenamefont {Kuklov},
  \citenamefont {Prokof'ev},\ and\ \citenamefont {Svistunov}}]{Chen2013}%
  \BibitemOpen
  \bibfield  {author} {\bibinfo {author} {\bibfnamefont {K.}~\bibnamefont
  {Chen}}, \bibinfo {author} {\bibfnamefont {Y.}~\bibnamefont {Huang}},
  \bibinfo {author} {\bibfnamefont {Y.}~\bibnamefont {Deng}}, \bibinfo {author}
  {\bibfnamefont {A.}~\bibnamefont {Kuklov}}, \bibinfo {author} {\bibfnamefont
  {N.}~\bibnamefont {Prokof'ev}}, \ and\ \bibinfo {author} {\bibfnamefont
  {B.}~\bibnamefont {Svistunov}},\ }\href {\doibase
  10.1103/PhysRevLett.110.185701} {\bibfield  {journal} {\bibinfo  {journal}
  {Physical review letters}\ }\textbf {\bibinfo {volume} {110}},\ \bibinfo
  {pages} {185701} (\bibinfo {year} {2013})}\BibitemShut {NoStop}%
\bibitem [{\citenamefont {Block}\ \emph {et~al.}(2013)\citenamefont {Block},
  \citenamefont {Melko},\ and\ \citenamefont {Kaul}}]{Block2013}%
  \BibitemOpen
  \bibfield  {author} {\bibinfo {author} {\bibfnamefont {M.~S.}\ \bibnamefont
  {Block}}, \bibinfo {author} {\bibfnamefont {R.~G.}\ \bibnamefont {Melko}}, \
  and\ \bibinfo {author} {\bibfnamefont {R.~K.}\ \bibnamefont {Kaul}},\ }\href
  {\doibase 10.1103/PhysRevLett.111.137202} {\bibfield  {journal} {\bibinfo
  {journal} {Physical review letters}\ }\textbf {\bibinfo {volume} {111}},\
  \bibinfo {pages} {137202} (\bibinfo {year} {2013})}\BibitemShut {NoStop}%
\bibitem [{\citenamefont {Sreejith}\ and\ \citenamefont
  {Powell}(2015)}]{Sreejith2015}%
  \BibitemOpen
  \bibfield  {author} {\bibinfo {author} {\bibfnamefont {G.}~\bibnamefont
  {Sreejith}}\ and\ \bibinfo {author} {\bibfnamefont {S.}~\bibnamefont
  {Powell}},\ }\href {\doibase 10.1103/PhysRevB.92.184413} {\bibfield
  {journal} {\bibinfo  {journal} {Physical Review B}\ }\textbf {\bibinfo
  {volume} {92}},\ \bibinfo {pages} {184413} (\bibinfo {year}
  {2015})}\BibitemShut {NoStop}%
\bibitem [{\citenamefont {Nahum}\ \emph
  {et~al.}(2015{\natexlab{a}})\citenamefont {Nahum}, \citenamefont {Chalker},
  \citenamefont {Serna}, \citenamefont {Ortu{\~n}o},\ and\ \citenamefont
  {Somoza}}]{Nahum2015}%
  \BibitemOpen
  \bibfield  {author} {\bibinfo {author} {\bibfnamefont {A.}~\bibnamefont
  {Nahum}}, \bibinfo {author} {\bibfnamefont {J.}~\bibnamefont {Chalker}},
  \bibinfo {author} {\bibfnamefont {P.}~\bibnamefont {Serna}}, \bibinfo
  {author} {\bibfnamefont {M.}~\bibnamefont {Ortu{\~n}o}}, \ and\ \bibinfo
  {author} {\bibfnamefont {A.}~\bibnamefont {Somoza}},\ }\href {\doibase
  10.1103/PhysRevX.5.041048} {\bibfield  {journal} {\bibinfo  {journal}
  {Physical Review X}\ }\textbf {\bibinfo {volume} {5}},\ \bibinfo {pages}
  {041048} (\bibinfo {year} {2015}{\natexlab{a}})}\BibitemShut {NoStop}%
\bibitem [{\citenamefont {Nahum}\ \emph
  {et~al.}(2015{\natexlab{b}})\citenamefont {Nahum}, \citenamefont {Serna},
  \citenamefont {Chalker}, \citenamefont {Ortu{\~n}o},\ and\ \citenamefont
  {Somoza}}]{Nahum2015PRL}%
  \BibitemOpen
  \bibfield  {author} {\bibinfo {author} {\bibfnamefont {A.}~\bibnamefont
  {Nahum}}, \bibinfo {author} {\bibfnamefont {P.}~\bibnamefont {Serna}},
  \bibinfo {author} {\bibfnamefont {J.}~\bibnamefont {Chalker}}, \bibinfo
  {author} {\bibfnamefont {M.}~\bibnamefont {Ortu{\~n}o}}, \ and\ \bibinfo
  {author} {\bibfnamefont {A.}~\bibnamefont {Somoza}},\ }\href {\doibase
  10.1103/PhysRevLett.115.267203} {\bibfield  {journal} {\bibinfo  {journal}
  {Physical review letters}\ }\textbf {\bibinfo {volume} {115}},\ \bibinfo
  {pages} {267203} (\bibinfo {year} {2015}{\natexlab{b}})}\BibitemShut
  {NoStop}%
\bibitem [{\citenamefont {Shao}\ \emph {et~al.}(2016)\citenamefont {Shao},
  \citenamefont {Guo},\ and\ \citenamefont {Sandvik}}]{Shao2016}%
  \BibitemOpen
  \bibfield  {author} {\bibinfo {author} {\bibfnamefont {H.}~\bibnamefont
  {Shao}}, \bibinfo {author} {\bibfnamefont {W.}~\bibnamefont {Guo}}, \ and\
  \bibinfo {author} {\bibfnamefont {A.~W.}\ \bibnamefont {Sandvik}},\ }\href
  {\doibase 10.1126/science.aad5007} {\bibfield  {journal} {\bibinfo  {journal}
  {Science}\ }\textbf {\bibinfo {volume} {352}},\ \bibinfo {pages} {213}
  (\bibinfo {year} {2016})}\BibitemShut {NoStop}%
\bibitem [{\citenamefont {Shao}\ \emph {et~al.}(2017)\citenamefont {Shao},
  \citenamefont {Qin}, \citenamefont {Capponi}, \citenamefont {Chesi},
  \citenamefont {Meng},\ and\ \citenamefont {Sandvik}}]{Shao2017}%
  \BibitemOpen
  \bibfield  {author} {\bibinfo {author} {\bibfnamefont {H.}~\bibnamefont
  {Shao}}, \bibinfo {author} {\bibfnamefont {Y.~Q.}\ \bibnamefont {Qin}},
  \bibinfo {author} {\bibfnamefont {S.}~\bibnamefont {Capponi}}, \bibinfo
  {author} {\bibfnamefont {S.}~\bibnamefont {Chesi}}, \bibinfo {author}
  {\bibfnamefont {Z.~Y.}\ \bibnamefont {Meng}}, \ and\ \bibinfo {author}
  {\bibfnamefont {A.~W.}\ \bibnamefont {Sandvik}},\ }\href {\doibase
  10.1103/PhysRevX.7.041072} {\bibfield  {journal} {\bibinfo  {journal}
  {Physical Review X}\ }\textbf {\bibinfo {volume} {7}},\ \bibinfo {pages}
  {041072} (\bibinfo {year} {2017})}\BibitemShut {NoStop}%
\bibitem [{\citenamefont {Qin}\ \emph {et~al.}(2017)\citenamefont {Qin},
  \citenamefont {He}, \citenamefont {You}, \citenamefont {Lu}, \citenamefont
  {Sen}, \citenamefont {Sandvik}, \citenamefont {Xu},\ and\ \citenamefont
  {Meng}}]{Qin2017}%
  \BibitemOpen
  \bibfield  {author} {\bibinfo {author} {\bibfnamefont {Y.~Q.}\ \bibnamefont
  {Qin}}, \bibinfo {author} {\bibfnamefont {Y.-Y.}\ \bibnamefont {He}},
  \bibinfo {author} {\bibfnamefont {Y.-Z.}\ \bibnamefont {You}}, \bibinfo
  {author} {\bibfnamefont {Z.-Y.}\ \bibnamefont {Lu}}, \bibinfo {author}
  {\bibfnamefont {A.}~\bibnamefont {Sen}}, \bibinfo {author} {\bibfnamefont
  {A.~W.}\ \bibnamefont {Sandvik}}, \bibinfo {author} {\bibfnamefont
  {C.}~\bibnamefont {Xu}}, \ and\ \bibinfo {author} {\bibfnamefont {Z.~Y.}\
  \bibnamefont {Meng}},\ }\href {\doibase 10.1103/PhysRevX.7.031052} {\bibfield
   {journal} {\bibinfo  {journal} {Physical Review X}\ }\textbf {\bibinfo
  {volume} {7}},\ \bibinfo {pages} {031052} (\bibinfo {year}
  {2017})}\BibitemShut {NoStop}%
\bibitem [{\citenamefont {Sato}\ \emph {et~al.}(2017)\citenamefont {Sato},
  \citenamefont {Hohenadler},\ and\ \citenamefont {Assaad}}]{Sato2017}%
  \BibitemOpen
  \bibfield  {author} {\bibinfo {author} {\bibfnamefont {T.}~\bibnamefont
  {Sato}}, \bibinfo {author} {\bibfnamefont {M.}~\bibnamefont {Hohenadler}}, \
  and\ \bibinfo {author} {\bibfnamefont {F.~F.}\ \bibnamefont {Assaad}},\
  }\href {\doibase 10.1103/PhysRevLett.119.197203} {\bibfield  {journal}
  {\bibinfo  {journal} {Physical review letters}\ }\textbf {\bibinfo {volume}
  {119}},\ \bibinfo {pages} {197203} (\bibinfo {year} {2017})}\BibitemShut
  {NoStop}%
\bibitem [{\citenamefont {Ma}\ \emph {et~al.}(2018)\citenamefont {Ma},
  \citenamefont {Sun}, \citenamefont {You}, \citenamefont {Xu}, \citenamefont
  {Vishwanath}, \citenamefont {Sandvik},\ and\ \citenamefont {Meng}}]{Ma2018}%
  \BibitemOpen
  \bibfield  {author} {\bibinfo {author} {\bibfnamefont {N.}~\bibnamefont
  {Ma}}, \bibinfo {author} {\bibfnamefont {G.-Y.}\ \bibnamefont {Sun}},
  \bibinfo {author} {\bibfnamefont {Y.-Z.}\ \bibnamefont {You}}, \bibinfo
  {author} {\bibfnamefont {C.}~\bibnamefont {Xu}}, \bibinfo {author}
  {\bibfnamefont {A.}~\bibnamefont {Vishwanath}}, \bibinfo {author}
  {\bibfnamefont {A.~W.}\ \bibnamefont {Sandvik}}, \ and\ \bibinfo {author}
  {\bibfnamefont {Z.~Y.}\ \bibnamefont {Meng}},\ }\href {\doibase
  10.1103/PhysRevB.98.174421} {\bibfield  {journal} {\bibinfo  {journal}
  {Physical Review B}\ }\textbf {\bibinfo {volume} {98}},\ \bibinfo {pages}
  {174421} (\bibinfo {year} {2018})}\BibitemShut {NoStop}%
\bibitem [{\citenamefont {Zhao}\ \emph {et~al.}(2018)\citenamefont {Zhao},
  \citenamefont {Weinberg},\ and\ \citenamefont {Sandvik}}]{Zhao2018}%
  \BibitemOpen
  \bibfield  {author} {\bibinfo {author} {\bibfnamefont {B.}~\bibnamefont
  {Zhao}}, \bibinfo {author} {\bibfnamefont {P.}~\bibnamefont {Weinberg}}, \
  and\ \bibinfo {author} {\bibfnamefont {A.~W.}\ \bibnamefont {Sandvik}},\
  }\href {https://arxiv.org/abs/1804.07115} {\bibfield  {journal} {\bibinfo
  {journal} {arXiv preprint arXiv:1804.07115}\ } (\bibinfo {year}
  {2018})}\BibitemShut {NoStop}%
\bibitem [{\citenamefont {Serna}\ and\ \citenamefont
  {Nahum}(2019)}]{Serna2019}%
  \BibitemOpen
  \bibfield  {author} {\bibinfo {author} {\bibfnamefont {P.}~\bibnamefont
  {Serna}}\ and\ \bibinfo {author} {\bibfnamefont {A.}~\bibnamefont {Nahum}},\
  }\href {\doibase 10.1103/PhysRevB.99.195110} {\bibfield  {journal} {\bibinfo
  {journal} {Physical Review B}\ }\textbf {\bibinfo {volume} {99}},\ \bibinfo
  {pages} {195110} (\bibinfo {year} {2019})}\BibitemShut {NoStop}%
\bibitem [{\citenamefont {Ippoliti}\ \emph {et~al.}(2018)\citenamefont
  {Ippoliti}, \citenamefont {Mong}, \citenamefont {Assaad},\ and\ \citenamefont
  {Zaletel}}]{Ippoliti2018}%
  \BibitemOpen
  \bibfield  {author} {\bibinfo {author} {\bibfnamefont {M.}~\bibnamefont
  {Ippoliti}}, \bibinfo {author} {\bibfnamefont {R.~S.}\ \bibnamefont {Mong}},
  \bibinfo {author} {\bibfnamefont {F.~F.}\ \bibnamefont {Assaad}}, \ and\
  \bibinfo {author} {\bibfnamefont {M.~P.}\ \bibnamefont {Zaletel}},\ }\href
  {\doibase 10.1103/PhysRevB.98.235108} {\bibfield  {journal} {\bibinfo
  {journal} {Physical Review B}\ }\textbf {\bibinfo {volume} {98}},\ \bibinfo
  {pages} {235108} (\bibinfo {year} {2018})}\BibitemShut {NoStop}%
\bibitem [{\citenamefont {Zhang}\ \emph {et~al.}(2018)\citenamefont {Zhang},
  \citenamefont {He}, \citenamefont {Eggert}, \citenamefont {Moessner},\ and\
  \citenamefont {Pollmann}}]{Zhang2018}%
  \BibitemOpen
  \bibfield  {author} {\bibinfo {author} {\bibfnamefont {X.-F.}\ \bibnamefont
  {Zhang}}, \bibinfo {author} {\bibfnamefont {Y.-C.}\ \bibnamefont {He}},
  \bibinfo {author} {\bibfnamefont {S.}~\bibnamefont {Eggert}}, \bibinfo
  {author} {\bibfnamefont {R.}~\bibnamefont {Moessner}}, \ and\ \bibinfo
  {author} {\bibfnamefont {F.}~\bibnamefont {Pollmann}},\ }\href {\doibase
  10.1103/PhysRevLett.120.115702} {\bibfield  {journal} {\bibinfo  {journal}
  {Physical review letters}\ }\textbf {\bibinfo {volume} {120}},\ \bibinfo
  {pages} {115702} (\bibinfo {year} {2018})}\BibitemShut {NoStop}%
\bibitem [{\citenamefont {Patil}\ \emph {et~al.}(2018)\citenamefont {Patil},
  \citenamefont {Katz},\ and\ \citenamefont {Sandvik}}]{Patil2018}%
  \BibitemOpen
  \bibfield  {author} {\bibinfo {author} {\bibfnamefont {P.}~\bibnamefont
  {Patil}}, \bibinfo {author} {\bibfnamefont {E.}~\bibnamefont {Katz}}, \ and\
  \bibinfo {author} {\bibfnamefont {A.~W.}\ \bibnamefont {Sandvik}},\ }\href
  {\doibase 10.1103/PhysRevB.98.014414} {\bibfield  {journal} {\bibinfo
  {journal} {Physical Review B}\ }\textbf {\bibinfo {volume} {98}},\ \bibinfo
  {pages} {014414} (\bibinfo {year} {2018})}\BibitemShut {NoStop}%
\bibitem [{\citenamefont {Jiang}\ and\ \citenamefont
  {Motrunich}(2019)}]{Jiang2019}%
  \BibitemOpen
  \bibfield  {author} {\bibinfo {author} {\bibfnamefont {S.}~\bibnamefont
  {Jiang}}\ and\ \bibinfo {author} {\bibfnamefont {O.}~\bibnamefont
  {Motrunich}},\ }\href {\doibase 10.1103/PhysRevB.99.075103} {\bibfield
  {journal} {\bibinfo  {journal} {Physical Review B}\ }\textbf {\bibinfo
  {volume} {99}},\ \bibinfo {pages} {075103} (\bibinfo {year}
  {2019})}\BibitemShut {NoStop}%
\bibitem [{\citenamefont {Mudry}\ \emph {et~al.}(2019)\citenamefont {Mudry},
  \citenamefont {Furusaki}, \citenamefont {Morimoto},\ and\ \citenamefont
  {Hikihara}}]{Mudry2019}%
  \BibitemOpen
  \bibfield  {author} {\bibinfo {author} {\bibfnamefont {C.}~\bibnamefont
  {Mudry}}, \bibinfo {author} {\bibfnamefont {A.}~\bibnamefont {Furusaki}},
  \bibinfo {author} {\bibfnamefont {T.}~\bibnamefont {Morimoto}}, \ and\
  \bibinfo {author} {\bibfnamefont {T.}~\bibnamefont {Hikihara}},\ }\href
  {https://arxiv.org/abs/1903.05646} {\bibfield  {journal} {\bibinfo  {journal}
  {arXiv preprint arXiv:1903.05646}\ } (\bibinfo {year} {2019})}\BibitemShut
  {NoStop}%
\bibitem [{\citenamefont {Roberts}\ \emph {et~al.}(2019)\citenamefont
  {Roberts}, \citenamefont {Jiang},\ and\ \citenamefont
  {Motrunich}}]{Roberts2019}%
  \BibitemOpen
  \bibfield  {author} {\bibinfo {author} {\bibfnamefont {B.}~\bibnamefont
  {Roberts}}, \bibinfo {author} {\bibfnamefont {S.}~\bibnamefont {Jiang}}, \
  and\ \bibinfo {author} {\bibfnamefont {O.~I.}\ \bibnamefont {Motrunich}},\
  }\href {\doibase 10.1103/PhysRevB.99.165143} {\bibfield  {journal} {\bibinfo
  {journal} {Physical Review B}\ }\textbf {\bibinfo {volume} {99}},\ \bibinfo
  {pages} {165143} (\bibinfo {year} {2019})}\BibitemShut {NoStop}%
\bibitem [{\citenamefont {Huang}\ \emph {et~al.}(2019)\citenamefont {Huang},
  \citenamefont {Lu}, \citenamefont {You}, \citenamefont {Meng},\ and\
  \citenamefont {Xiang}}]{Huang2019}%
  \BibitemOpen
  \bibfield  {author} {\bibinfo {author} {\bibfnamefont {R.-Z.}\ \bibnamefont
  {Huang}}, \bibinfo {author} {\bibfnamefont {D.-C.}\ \bibnamefont {Lu}},
  \bibinfo {author} {\bibfnamefont {Y.-Z.}\ \bibnamefont {You}}, \bibinfo
  {author} {\bibfnamefont {Z.~Y.}\ \bibnamefont {Meng}}, \ and\ \bibinfo
  {author} {\bibfnamefont {T.}~\bibnamefont {Xiang}},\ }\href
  {https://arxiv.org/abs/1904.00021} {\bibfield  {journal} {\bibinfo  {journal}
  {arXiv preprint arXiv:1904.00021}\ } (\bibinfo {year} {2019})}\BibitemShut
  {NoStop}%
\bibitem [{\citenamefont {Affleck}\ and\ \citenamefont
  {Haldane}(1987)}]{Affleck1987}%
  \BibitemOpen
  \bibfield  {author} {\bibinfo {author} {\bibfnamefont {I.}~\bibnamefont
  {Affleck}}\ and\ \bibinfo {author} {\bibfnamefont {F.}~\bibnamefont
  {Haldane}},\ }\href {\doibase 10.1103/PhysRevB.36.5291} {\bibfield  {journal}
  {\bibinfo  {journal} {Physical Review B}\ }\textbf {\bibinfo {volume} {36}},\
  \bibinfo {pages} {5291} (\bibinfo {year} {1987})}\BibitemShut {NoStop}%
\bibitem [{\citenamefont {Sengupta}\ \emph {et~al.}(2002)\citenamefont
  {Sengupta}, \citenamefont {Sandvik},\ and\ \citenamefont
  {Campbell}}]{Sengupta2002}%
  \BibitemOpen
  \bibfield  {author} {\bibinfo {author} {\bibfnamefont {P.}~\bibnamefont
  {Sengupta}}, \bibinfo {author} {\bibfnamefont {A.~W.}\ \bibnamefont
  {Sandvik}}, \ and\ \bibinfo {author} {\bibfnamefont {D.~K.}\ \bibnamefont
  {Campbell}},\ }\href {\doibase 10.1103/PhysRevB.65.155113} {\bibfield
  {journal} {\bibinfo  {journal} {Physical Review B}\ }\textbf {\bibinfo
  {volume} {65}},\ \bibinfo {pages} {155113} (\bibinfo {year}
  {2002})}\BibitemShut {NoStop}%
\bibitem [{\citenamefont {Sandvik}\ \emph {et~al.}(2004)\citenamefont
  {Sandvik}, \citenamefont {Balents},\ and\ \citenamefont
  {Campbell}}]{Sandvik2004}%
  \BibitemOpen
  \bibfield  {author} {\bibinfo {author} {\bibfnamefont {A.~W.}\ \bibnamefont
  {Sandvik}}, \bibinfo {author} {\bibfnamefont {L.}~\bibnamefont {Balents}}, \
  and\ \bibinfo {author} {\bibfnamefont {D.~K.}\ \bibnamefont {Campbell}},\
  }\href {\doibase 10.1103/PhysRevLett.92.236401} {\bibfield  {journal}
  {\bibinfo  {journal} {Physical review letters}\ }\textbf {\bibinfo {volume}
  {92}},\ \bibinfo {pages} {236401} (\bibinfo {year} {2004})}\BibitemShut
  {NoStop}%
\bibitem [{\citenamefont {White}(1992)}]{White1992}%
  \BibitemOpen
  \bibfield  {author} {\bibinfo {author} {\bibfnamefont {S.~R.}\ \bibnamefont
  {White}},\ }\href {\doibase 10.1103/PhysRevLett.69.2863} {\bibfield
  {journal} {\bibinfo  {journal} {Physical review letters}\ }\textbf {\bibinfo
  {volume} {69}},\ \bibinfo {pages} {2863} (\bibinfo {year}
  {1992})}\BibitemShut {NoStop}%
\bibitem [{\citenamefont {Schollw{\"o}ck}(2005)}]{Schollwock2005}%
  \BibitemOpen
  \bibfield  {author} {\bibinfo {author} {\bibfnamefont {U.}~\bibnamefont
  {Schollw{\"o}ck}},\ }\href {\doibase 10.1103/RevModPhys.77.259} {\bibfield
  {journal} {\bibinfo  {journal} {Reviews of modern physics}\ }\textbf
  {\bibinfo {volume} {77}},\ \bibinfo {pages} {259} (\bibinfo {year}
  {2005})}\BibitemShut {NoStop}%
\bibitem [{\citenamefont {Verstraete}\ \emph {et~al.}(2004)\citenamefont
  {Verstraete}, \citenamefont {Porras},\ and\ \citenamefont
  {Cirac}}]{Verstraete2004}%
  \BibitemOpen
  \bibfield  {author} {\bibinfo {author} {\bibfnamefont {F.}~\bibnamefont
  {Verstraete}}, \bibinfo {author} {\bibfnamefont {D.}~\bibnamefont {Porras}},
  \ and\ \bibinfo {author} {\bibfnamefont {J.~I.}\ \bibnamefont {Cirac}},\
  }\href {\doibase 10.1103/PhysRevLett.93.227205} {\bibfield  {journal}
  {\bibinfo  {journal} {Physical review letters}\ }\textbf {\bibinfo {volume}
  {93}},\ \bibinfo {pages} {227205} (\bibinfo {year} {2004})}\BibitemShut
  {NoStop}%
\bibitem [{\citenamefont {Schollw{\"o}ck}(2011)}]{Schollwock2011}%
  \BibitemOpen
  \bibfield  {author} {\bibinfo {author} {\bibfnamefont {U.}~\bibnamefont
  {Schollw{\"o}ck}},\ }\href {\doibase 10.1016/j.aop.2010.09.012} {\bibfield
  {journal} {\bibinfo  {journal} {Annals of Physics}\ }\textbf {\bibinfo
  {volume} {326}},\ \bibinfo {pages} {96} (\bibinfo {year} {2011})}\BibitemShut
  {NoStop}%
\bibitem [{\citenamefont {Gu}(2010)}]{Gu2010}%
  \BibitemOpen
  \bibfield  {author} {\bibinfo {author} {\bibfnamefont {S.-J.}\ \bibnamefont
  {Gu}},\ }\href {\doibase 10.1142/S0217979210056335} {\bibfield  {journal}
  {\bibinfo  {journal} {International Journal of Modern Physics B}\ }\textbf
  {\bibinfo {volume} {24}},\ \bibinfo {pages} {4371} (\bibinfo {year}
  {2010})}\BibitemShut {NoStop}%
\bibitem [{\citenamefont {Zanardi}\ and\ \citenamefont
  {Paunkovi\ifmmode~\acute{c}\else \'{c}\fi{}}(2006)}]{Zanardi2006}%
  \BibitemOpen
  \bibfield  {author} {\bibinfo {author} {\bibfnamefont {P.}~\bibnamefont
  {Zanardi}}\ and\ \bibinfo {author} {\bibfnamefont {N.}~\bibnamefont
  {Paunkovi\ifmmode~\acute{c}\else \'{c}\fi{}}},\ }\href {\doibase
  10.1103/PhysRevE.74.031123} {\bibfield  {journal} {\bibinfo  {journal} {Phys.
  Rev. E}\ }\textbf {\bibinfo {volume} {74}},\ \bibinfo {pages} {031123}
  (\bibinfo {year} {2006})}\BibitemShut {NoStop}%
\bibitem [{\citenamefont {You}\ \emph {et~al.}(2007)\citenamefont {You},
  \citenamefont {Li},\ and\ \citenamefont {Gu}}]{You2007}%
  \BibitemOpen
  \bibfield  {author} {\bibinfo {author} {\bibfnamefont {W.-L.}\ \bibnamefont
  {You}}, \bibinfo {author} {\bibfnamefont {Y.-W.}\ \bibnamefont {Li}}, \ and\
  \bibinfo {author} {\bibfnamefont {S.-J.}\ \bibnamefont {Gu}},\ }\href
  {\doibase 10.1103/PhysRevE.76.022101} {\bibfield  {journal} {\bibinfo
  {journal} {Physical Review E}\ }\textbf {\bibinfo {volume} {76}},\ \bibinfo
  {pages} {022101} (\bibinfo {year} {2007})}\BibitemShut {NoStop}%
\bibitem [{\citenamefont {Venuti}\ and\ \citenamefont
  {Zanardi}(2007)}]{Venuti2007}%
  \BibitemOpen
  \bibfield  {author} {\bibinfo {author} {\bibfnamefont {L.~C.}\ \bibnamefont
  {Venuti}}\ and\ \bibinfo {author} {\bibfnamefont {P.}~\bibnamefont
  {Zanardi}},\ }\href {\doibase 10.1103/PhysRevLett.99.095701} {\bibfield
  {journal} {\bibinfo  {journal} {Physical review letters}\ }\textbf {\bibinfo
  {volume} {99}},\ \bibinfo {pages} {095701} (\bibinfo {year}
  {2007})}\BibitemShut {NoStop}%
\bibitem [{\citenamefont {Chen}\ \emph {et~al.}(2008)\citenamefont {Chen},
  \citenamefont {Wang}, \citenamefont {Hao},\ and\ \citenamefont
  {Wang}}]{Chen2008}%
  \BibitemOpen
  \bibfield  {author} {\bibinfo {author} {\bibfnamefont {S.}~\bibnamefont
  {Chen}}, \bibinfo {author} {\bibfnamefont {L.}~\bibnamefont {Wang}}, \bibinfo
  {author} {\bibfnamefont {Y.}~\bibnamefont {Hao}}, \ and\ \bibinfo {author}
  {\bibfnamefont {Y.}~\bibnamefont {Wang}},\ }\href {\doibase
  10.1103/PhysRevA.77.032111} {\bibfield  {journal} {\bibinfo  {journal}
  {Physical Review A}\ }\textbf {\bibinfo {volume} {77}},\ \bibinfo {pages}
  {032111} (\bibinfo {year} {2008})}\BibitemShut {NoStop}%
\bibitem [{\citenamefont {Gu}\ \emph {et~al.}(2008)\citenamefont {Gu},
  \citenamefont {Kwok}, \citenamefont {Ning},\ and\ \citenamefont
  {Lin}}]{Gu2008}%
  \BibitemOpen
  \bibfield  {author} {\bibinfo {author} {\bibfnamefont {S.-J.}\ \bibnamefont
  {Gu}}, \bibinfo {author} {\bibfnamefont {H.-M.}\ \bibnamefont {Kwok}},
  \bibinfo {author} {\bibfnamefont {W.-Q.}\ \bibnamefont {Ning}}, \ and\
  \bibinfo {author} {\bibfnamefont {H.-Q.}\ \bibnamefont {Lin}},\ }\href
  {\doibase 10.1103/PhysRevB.77.245109} {\bibfield  {journal} {\bibinfo
  {journal} {Phys. Rev. B}\ }\textbf {\bibinfo {volume} {77}},\ \bibinfo
  {pages} {245109} (\bibinfo {year} {2008})}\BibitemShut {NoStop}%
\bibitem [{\citenamefont {Yang}\ \emph {et~al.}(2008)\citenamefont {Yang},
  \citenamefont {Gu}, \citenamefont {Sun},\ and\ \citenamefont
  {Lin}}]{Yang2008}%
  \BibitemOpen
  \bibfield  {author} {\bibinfo {author} {\bibfnamefont {S.}~\bibnamefont
  {Yang}}, \bibinfo {author} {\bibfnamefont {S.-J.}\ \bibnamefont {Gu}},
  \bibinfo {author} {\bibfnamefont {C.-P.}\ \bibnamefont {Sun}}, \ and\
  \bibinfo {author} {\bibfnamefont {H.-Q.}\ \bibnamefont {Lin}},\ }\href
  {\doibase 10.1103/PhysRevA.78.012304} {\bibfield  {journal} {\bibinfo
  {journal} {Phys. Rev. A}\ }\textbf {\bibinfo {volume} {78}},\ \bibinfo
  {pages} {012304} (\bibinfo {year} {2008})}\BibitemShut {NoStop}%
\bibitem [{\citenamefont {Kwok}\ \emph {et~al.}(2008)\citenamefont {Kwok},
  \citenamefont {Ning}, \citenamefont {Gu},\ and\ \citenamefont
  {Lin}}]{Kwok2008}%
  \BibitemOpen
  \bibfield  {author} {\bibinfo {author} {\bibfnamefont {H.-M.}\ \bibnamefont
  {Kwok}}, \bibinfo {author} {\bibfnamefont {W.-Q.}\ \bibnamefont {Ning}},
  \bibinfo {author} {\bibfnamefont {S.-J.}\ \bibnamefont {Gu}}, \ and\ \bibinfo
  {author} {\bibfnamefont {H.-Q.}\ \bibnamefont {Lin}},\ }\href {\doibase
  10.1103/PhysRevE.78.032103} {\bibfield  {journal} {\bibinfo  {journal} {Phys.
  Rev. E}\ }\textbf {\bibinfo {volume} {78}},\ \bibinfo {pages} {032103}
  (\bibinfo {year} {2008})}\BibitemShut {NoStop}%
\bibitem [{\citenamefont {Gong}\ and\ \citenamefont {Tong}(2008)}]{Gong2008}%
  \BibitemOpen
  \bibfield  {author} {\bibinfo {author} {\bibfnamefont {L.}~\bibnamefont
  {Gong}}\ and\ \bibinfo {author} {\bibfnamefont {P.}~\bibnamefont {Tong}},\
  }\href {\doibase 10.1103/PhysRevB.78.115114} {\bibfield  {journal} {\bibinfo
  {journal} {Phys. Rev. B}\ }\textbf {\bibinfo {volume} {78}},\ \bibinfo
  {pages} {115114} (\bibinfo {year} {2008})}\BibitemShut {NoStop}%
\bibitem [{\citenamefont {Yu}\ \emph {et~al.}(2009)\citenamefont {Yu},
  \citenamefont {Kwok}, \citenamefont {Cao},\ and\ \citenamefont
  {Gu}}]{Yu2009}%
  \BibitemOpen
  \bibfield  {author} {\bibinfo {author} {\bibfnamefont {W.-C.}\ \bibnamefont
  {Yu}}, \bibinfo {author} {\bibfnamefont {H.-M.}\ \bibnamefont {Kwok}},
  \bibinfo {author} {\bibfnamefont {J.}~\bibnamefont {Cao}}, \ and\ \bibinfo
  {author} {\bibfnamefont {S.-J.}\ \bibnamefont {Gu}},\ }\href {\doibase
  10.1103/PhysRevE.80.021108} {\bibfield  {journal} {\bibinfo  {journal} {Phys.
  Rev. E}\ }\textbf {\bibinfo {volume} {80}},\ \bibinfo {pages} {021108}
  (\bibinfo {year} {2009})}\BibitemShut {NoStop}%
\bibitem [{\citenamefont {Schwandt}\ \emph {et~al.}(2009)\citenamefont
  {Schwandt}, \citenamefont {Alet},\ and\ \citenamefont
  {Capponi}}]{Schwandt2009}%
  \BibitemOpen
  \bibfield  {author} {\bibinfo {author} {\bibfnamefont {D.}~\bibnamefont
  {Schwandt}}, \bibinfo {author} {\bibfnamefont {F.}~\bibnamefont {Alet}}, \
  and\ \bibinfo {author} {\bibfnamefont {S.}~\bibnamefont {Capponi}},\ }\href
  {\doibase 10.1103/PhysRevLett.103.170501} {\bibfield  {journal} {\bibinfo
  {journal} {Phys. Rev. Lett.}\ }\textbf {\bibinfo {volume} {103}},\ \bibinfo
  {pages} {170501} (\bibinfo {year} {2009})}\BibitemShut {NoStop}%
\bibitem [{\citenamefont {Albuquerque}\ \emph {et~al.}(2010)\citenamefont
  {Albuquerque}, \citenamefont {Alet}, \citenamefont {Sire},\ and\
  \citenamefont {Capponi}}]{Albuquerque2010}%
  \BibitemOpen
  \bibfield  {author} {\bibinfo {author} {\bibfnamefont {A.~F.}\ \bibnamefont
  {Albuquerque}}, \bibinfo {author} {\bibfnamefont {F.}~\bibnamefont {Alet}},
  \bibinfo {author} {\bibfnamefont {C.}~\bibnamefont {Sire}}, \ and\ \bibinfo
  {author} {\bibfnamefont {S.}~\bibnamefont {Capponi}},\ }\href {\doibase
  10.1103/PhysRevB.81.064418} {\bibfield  {journal} {\bibinfo  {journal}
  {Physical Review B}\ }\textbf {\bibinfo {volume} {81}},\ \bibinfo {pages}
  {064418} (\bibinfo {year} {2010})}\BibitemShut {NoStop}%
\bibitem [{\citenamefont {Rams}\ and\ \citenamefont {Damski}(2011)}]{Rams2011}%
  \BibitemOpen
  \bibfield  {author} {\bibinfo {author} {\bibfnamefont {M.~M.}\ \bibnamefont
  {Rams}}\ and\ \bibinfo {author} {\bibfnamefont {B.}~\bibnamefont {Damski}},\
  }\href {\doibase 10.1103/PhysRevLett.106.055701} {\bibfield  {journal}
  {\bibinfo  {journal} {Phys. Rev. Lett.}\ }\textbf {\bibinfo {volume} {106}},\
  \bibinfo {pages} {055701} (\bibinfo {year} {2011})}\BibitemShut {NoStop}%
\bibitem [{\citenamefont {Li}\ \emph {et~al.}(2012)\citenamefont {Li},
  \citenamefont {Shi}, \citenamefont {Su}, \citenamefont {Liu}, \citenamefont
  {Dai},\ and\ \citenamefont {Zhou}}]{Li2012}%
  \BibitemOpen
  \bibfield  {author} {\bibinfo {author} {\bibfnamefont {S.-H.}\ \bibnamefont
  {Li}}, \bibinfo {author} {\bibfnamefont {Q.-Q.}\ \bibnamefont {Shi}},
  \bibinfo {author} {\bibfnamefont {Y.-H.}\ \bibnamefont {Su}}, \bibinfo
  {author} {\bibfnamefont {J.-H.}\ \bibnamefont {Liu}}, \bibinfo {author}
  {\bibfnamefont {Y.-W.}\ \bibnamefont {Dai}}, \ and\ \bibinfo {author}
  {\bibfnamefont {H.-Q.}\ \bibnamefont {Zhou}},\ }\href {\doibase
  10.1103/PhysRevB.86.064401} {\bibfield  {journal} {\bibinfo  {journal} {Phys.
  Rev. B}\ }\textbf {\bibinfo {volume} {86}},\ \bibinfo {pages} {064401}
  (\bibinfo {year} {2012})}\BibitemShut {NoStop}%
\bibitem [{\citenamefont {Mukherjee}\ \emph {et~al.}(2012)\citenamefont
  {Mukherjee}, \citenamefont {Dutta},\ and\ \citenamefont {Sen}}]{Victor2012}%
  \BibitemOpen
  \bibfield  {author} {\bibinfo {author} {\bibfnamefont {V.}~\bibnamefont
  {Mukherjee}}, \bibinfo {author} {\bibfnamefont {A.}~\bibnamefont {Dutta}}, \
  and\ \bibinfo {author} {\bibfnamefont {D.}~\bibnamefont {Sen}},\ }\href
  {\doibase 10.1103/PhysRevB.85.024301} {\bibfield  {journal} {\bibinfo
  {journal} {Phys. Rev. B}\ }\textbf {\bibinfo {volume} {85}},\ \bibinfo
  {pages} {024301} (\bibinfo {year} {2012})}\BibitemShut {NoStop}%
\bibitem [{\citenamefont {Carrasquilla}\ \emph
  {et~al.}(2013{\natexlab{a}})\citenamefont {Carrasquilla}, \citenamefont
  {Manmana},\ and\ \citenamefont {Rigol}}]{Rigol2013}%
  \BibitemOpen
  \bibfield  {author} {\bibinfo {author} {\bibfnamefont {J.}~\bibnamefont
  {Carrasquilla}}, \bibinfo {author} {\bibfnamefont {S.~R.}\ \bibnamefont
  {Manmana}}, \ and\ \bibinfo {author} {\bibfnamefont {M.}~\bibnamefont
  {Rigol}},\ }\href {\doibase 10.1103/PhysRevA.87.043606} {\bibfield  {journal}
  {\bibinfo  {journal} {Phys. Rev. A}\ }\textbf {\bibinfo {volume} {87}},\
  \bibinfo {pages} {043606} (\bibinfo {year} {2013}{\natexlab{a}})}\BibitemShut
  {NoStop}%
\bibitem [{\citenamefont {Damski}(2013)}]{Damski2013}%
  \BibitemOpen
  \bibfield  {author} {\bibinfo {author} {\bibfnamefont {B.}~\bibnamefont
  {Damski}},\ }\href {\doibase 10.1103/PhysRevE.87.052131} {\bibfield
  {journal} {\bibinfo  {journal} {Phys. Rev. E}\ }\textbf {\bibinfo {volume}
  {87}},\ \bibinfo {pages} {052131} (\bibinfo {year} {2013})}\BibitemShut
  {NoStop}%
\bibitem [{\citenamefont {\L{}\k{a}cki}\ \emph {et~al.}(2014)\citenamefont
  {\L{}\k{a}cki}, \citenamefont {Damski},\ and\ \citenamefont
  {Zakrzewski}}]{Damski2014}%
  \BibitemOpen
  \bibfield  {author} {\bibinfo {author} {\bibfnamefont {M.}~\bibnamefont
  {\L{}\k{a}cki}}, \bibinfo {author} {\bibfnamefont {B.}~\bibnamefont
  {Damski}}, \ and\ \bibinfo {author} {\bibfnamefont {J.}~\bibnamefont
  {Zakrzewski}},\ }\href {\doibase 10.1103/PhysRevA.89.033625} {\bibfield
  {journal} {\bibinfo  {journal} {Phys. Rev. A}\ }\textbf {\bibinfo {volume}
  {89}},\ \bibinfo {pages} {033625} (\bibinfo {year} {2014})}\BibitemShut
  {NoStop}%
\bibitem [{\citenamefont {Sun}(2017)}]{Sun2017}%
  \BibitemOpen
  \bibfield  {author} {\bibinfo {author} {\bibfnamefont {G.}~\bibnamefont
  {Sun}},\ }\href {\doibase 10.1103/PhysRevA.96.043621} {\bibfield  {journal}
  {\bibinfo  {journal} {Physical Review A}\ }\textbf {\bibinfo {volume} {96}},\
  \bibinfo {pages} {043621} (\bibinfo {year} {2017})}\BibitemShut {NoStop}%
\bibitem [{\citenamefont {Wei}\ and\ \citenamefont {Lv}(2018)}]{Wei2018}%
  \BibitemOpen
  \bibfield  {author} {\bibinfo {author} {\bibfnamefont {B.-B.}\ \bibnamefont
  {Wei}}\ and\ \bibinfo {author} {\bibfnamefont {X.-C.}\ \bibnamefont {Lv}},\
  }\href {\doibase 10.1103/PhysRevA.97.013845} {\bibfield  {journal} {\bibinfo
  {journal} {Phys. Rev. A}\ }\textbf {\bibinfo {volume} {97}},\ \bibinfo
  {pages} {013845} (\bibinfo {year} {2018})}\BibitemShut {NoStop}%
\bibitem [{\citenamefont {Zhu}\ \emph {et~al.}(2018)\citenamefont {Zhu},
  \citenamefont {Sun}, \citenamefont {You},\ and\ \citenamefont
  {Shi}}]{Zhu2018}%
  \BibitemOpen
  \bibfield  {author} {\bibinfo {author} {\bibfnamefont {Z.}~\bibnamefont
  {Zhu}}, \bibinfo {author} {\bibfnamefont {G.}~\bibnamefont {Sun}}, \bibinfo
  {author} {\bibfnamefont {W.-L.}\ \bibnamefont {You}}, \ and\ \bibinfo
  {author} {\bibfnamefont {D.-N.}\ \bibnamefont {Shi}},\ }\href {\doibase
  10.1103/PhysRevA.98.023607} {\bibfield  {journal} {\bibinfo  {journal}
  {Physical Review A}\ }\textbf {\bibinfo {volume} {98}},\ \bibinfo {pages}
  {023607} (\bibinfo {year} {2018})}\BibitemShut {NoStop}%
\bibitem [{\citenamefont {Luo}\ \emph {et~al.}(2018)\citenamefont {Luo},
  \citenamefont {Zhao},\ and\ \citenamefont {Wang}}]{Lu2018}%
  \BibitemOpen
  \bibfield  {author} {\bibinfo {author} {\bibfnamefont {Q.}~\bibnamefont
  {Luo}}, \bibinfo {author} {\bibfnamefont {J.}~\bibnamefont {Zhao}}, \ and\
  \bibinfo {author} {\bibfnamefont {X.}~\bibnamefont {Wang}},\ }\href {\doibase
  10.1103/PhysRevE.98.022106} {\bibfield  {journal} {\bibinfo  {journal} {Phys.
  Rev. E}\ }\textbf {\bibinfo {volume} {98}},\ \bibinfo {pages} {022106}
  (\bibinfo {year} {2018})}\BibitemShut {NoStop}%
\bibitem [{\citenamefont {Wei}(2019)}]{Wei2019}%
  \BibitemOpen
  \bibfield  {author} {\bibinfo {author} {\bibfnamefont {B.-B.}\ \bibnamefont
  {Wei}},\ }\href {\doibase 10.1103/PhysRevA.99.042117} {\bibfield  {journal}
  {\bibinfo  {journal} {Physical Review A}\ }\textbf {\bibinfo {volume} {99}},\
  \bibinfo {pages} {042117} (\bibinfo {year} {2019})}\BibitemShut {NoStop}%
\bibitem [{\citenamefont {Yang}(2007)}]{Yang2007}%
  \BibitemOpen
  \bibfield  {author} {\bibinfo {author} {\bibfnamefont {M.-F.}\ \bibnamefont
  {Yang}},\ }\href {\doibase 10.1103/PhysRevB.76.180403} {\bibfield  {journal}
  {\bibinfo  {journal} {Physical Review B}\ }\textbf {\bibinfo {volume} {76}},\
  \bibinfo {pages} {180403} (\bibinfo {year} {2007})}\BibitemShut {NoStop}%
\bibitem [{\citenamefont {Fj{\ae}restad}(2008)}]{Fjestad2018}%
  \BibitemOpen
  \bibfield  {author} {\bibinfo {author} {\bibfnamefont {J.~O.}\ \bibnamefont
  {Fj{\ae}restad}},\ }\href {\doibase 10.1088/1742-5468/2008/07/P07011}
  {\bibfield  {journal} {\bibinfo  {journal} {Journal of Statistical Mechanics:
  Theory and Experiment}\ }\textbf {\bibinfo {volume} {2008}},\ \bibinfo
  {pages} {P07011} (\bibinfo {year} {2008})}\BibitemShut {NoStop}%
\bibitem [{\citenamefont {Langari}\ and\ \citenamefont
  {Rezakhani}(2012)}]{Langari2012}%
  \BibitemOpen
  \bibfield  {author} {\bibinfo {author} {\bibfnamefont {A.}~\bibnamefont
  {Langari}}\ and\ \bibinfo {author} {\bibfnamefont {A.}~\bibnamefont
  {Rezakhani}},\ }\href {\doibase 10.1088/1367-2630/14/5/053014} {\bibfield
  {journal} {\bibinfo  {journal} {New Journal of Physics}\ }\textbf {\bibinfo
  {volume} {14}},\ \bibinfo {pages} {053014} (\bibinfo {year}
  {2012})}\BibitemShut {NoStop}%
\bibitem [{\citenamefont {Carrasquilla}\ \emph
  {et~al.}(2013{\natexlab{b}})\citenamefont {Carrasquilla}, \citenamefont
  {Manmana},\ and\ \citenamefont {Rigol}}]{Carrasquilla2013}%
  \BibitemOpen
  \bibfield  {author} {\bibinfo {author} {\bibfnamefont {J.}~\bibnamefont
  {Carrasquilla}}, \bibinfo {author} {\bibfnamefont {S.~R.}\ \bibnamefont
  {Manmana}}, \ and\ \bibinfo {author} {\bibfnamefont {M.}~\bibnamefont
  {Rigol}},\ }\href {\doibase 10.1103/PhysRevA.87.043606} {\bibfield  {journal}
  {\bibinfo  {journal} {Physical Review A}\ }\textbf {\bibinfo {volume} {87}},\
  \bibinfo {pages} {043606} (\bibinfo {year} {2013}{\natexlab{b}})}\BibitemShut
  {NoStop}%
\bibitem [{\citenamefont {{\L}{\k{a}}cki}\ \emph {et~al.}(2014)\citenamefont
  {{\L}{\k{a}}cki}, \citenamefont {Damski},\ and\ \citenamefont
  {Zakrzewski}}]{Lacki2014}%
  \BibitemOpen
  \bibfield  {author} {\bibinfo {author} {\bibfnamefont {M.}~\bibnamefont
  {{\L}{\k{a}}cki}}, \bibinfo {author} {\bibfnamefont {B.}~\bibnamefont
  {Damski}}, \ and\ \bibinfo {author} {\bibfnamefont {J.}~\bibnamefont
  {Zakrzewski}},\ }\href {\doibase 10.1103/PhysRevA.89.033625} {\bibfield
  {journal} {\bibinfo  {journal} {Physical Review A}\ }\textbf {\bibinfo
  {volume} {89}},\ \bibinfo {pages} {033625} (\bibinfo {year}
  {2014})}\BibitemShut {NoStop}%
\bibitem [{\citenamefont {Sun}\ \emph {et~al.}(2015)\citenamefont {Sun},
  \citenamefont {Kolezhuk},\ and\ \citenamefont {Vekua}}]{Sun2015}%
  \BibitemOpen
  \bibfield  {author} {\bibinfo {author} {\bibfnamefont {G.}~\bibnamefont
  {Sun}}, \bibinfo {author} {\bibfnamefont {A.}~\bibnamefont {Kolezhuk}}, \
  and\ \bibinfo {author} {\bibfnamefont {T.}~\bibnamefont {Vekua}},\ }\href
  {\doibase 10.1103/PhysRevB.91.014418} {\bibfield  {journal} {\bibinfo
  {journal} {Physical Review B}\ }\textbf {\bibinfo {volume} {91}},\ \bibinfo
  {pages} {014418} (\bibinfo {year} {2015})}\BibitemShut {NoStop}%
\bibitem [{\citenamefont {Cincio}\ \emph {et~al.}(2019)\citenamefont {Cincio},
  \citenamefont {Rams}, \citenamefont {Dziarmaga},\ and\ \citenamefont
  {Zurek}}]{Cincio2019}%
  \BibitemOpen
  \bibfield  {author} {\bibinfo {author} {\bibfnamefont {L.}~\bibnamefont
  {Cincio}}, \bibinfo {author} {\bibfnamefont {M.~M.}\ \bibnamefont {Rams}},
  \bibinfo {author} {\bibfnamefont {J.}~\bibnamefont {Dziarmaga}}, \ and\
  \bibinfo {author} {\bibfnamefont {W.~H.}\ \bibnamefont {Zurek}},\ }\href
  {https://arxiv.org/abs/1906.05307} {\bibfield  {journal} {\bibinfo  {journal}
  {arXiv preprint arXiv:1906.05307}\ } (\bibinfo {year} {2019})}\BibitemShut
  {NoStop}%
\bibitem [{\citenamefont {Wang}\ \emph {et~al.}(2017)\citenamefont {Wang},
  \citenamefont {Nahum}, \citenamefont {Metlitski}, \citenamefont {Xu},\ and\
  \citenamefont {Senthil}}]{Wang2017PRX}%
  \BibitemOpen
  \bibfield  {author} {\bibinfo {author} {\bibfnamefont {C.}~\bibnamefont
  {Wang}}, \bibinfo {author} {\bibfnamefont {A.}~\bibnamefont {Nahum}},
  \bibinfo {author} {\bibfnamefont {M.~A.}\ \bibnamefont {Metlitski}}, \bibinfo
  {author} {\bibfnamefont {C.}~\bibnamefont {Xu}}, \ and\ \bibinfo {author}
  {\bibfnamefont {T.}~\bibnamefont {Senthil}},\ }\href
  {https://doi.org/10.1103/PhysRevX.7.031051} {\bibfield  {journal} {\bibinfo
  {journal} {Physical Review X 7, 031051}\ } (\bibinfo {year}
  {2017})}\BibitemShut {NoStop}%
\bibitem [{\citenamefont {Iino}\ \emph {et~al.}(2019)\citenamefont {Iino},
  \citenamefont {Morita}, \citenamefont {Kawashima},\ and\ \citenamefont
  {Sandvik}}]{Iino2019}%
  \BibitemOpen
  \bibfield  {author} {\bibinfo {author} {\bibfnamefont {S.}~\bibnamefont
  {Iino}}, \bibinfo {author} {\bibfnamefont {S.}~\bibnamefont {Morita}},
  \bibinfo {author} {\bibfnamefont {N.}~\bibnamefont {Kawashima}}, \ and\
  \bibinfo {author} {\bibfnamefont {A.~W.}\ \bibnamefont {Sandvik}},\ }\href
  {https://doi.org/10.7566/JPSJ.88.034006} {\bibfield  {journal} {\bibinfo
  {journal} {Journal of the Physical Society of Japan 88, 034006}\ } (\bibinfo
  {year} {2019})}\BibitemShut {NoStop}%
\bibitem [{\citenamefont {Hauke}\ \emph {et~al.}(2016)\citenamefont {Hauke},
  \citenamefont {Heyl}, \citenamefont {Tagliacozzo},\ and\ \citenamefont
  {Zoller}}]{Zoller2016}%
  \BibitemOpen
  \bibfield  {author} {\bibinfo {author} {\bibfnamefont {P.}~\bibnamefont
  {Hauke}}, \bibinfo {author} {\bibfnamefont {M.}~\bibnamefont {Heyl}},
  \bibinfo {author} {\bibfnamefont {L.}~\bibnamefont {Tagliacozzo}}, \ and\
  \bibinfo {author} {\bibfnamefont {P.}~\bibnamefont {Zoller}},\ }\href
  {https://doi.org/10.1038/nphys3700} {\bibfield  {journal} {\bibinfo
  {journal} {Nature Physics 12, 778}\ } (\bibinfo {year} {2016})}\BibitemShut
  {NoStop}%
\bibitem [{\citenamefont {Carollo}\ and\ \citenamefont
  {Pachos}(2005)}]{Carollo2005}%
  \BibitemOpen
  \bibfield  {author} {\bibinfo {author} {\bibfnamefont {A.~C.~M.}\
  \bibnamefont {Carollo}}\ and\ \bibinfo {author} {\bibfnamefont {J.~K.}\
  \bibnamefont {Pachos}},\ }\href {\doibase 10.1103/PhysRevLett.95.157203}
  {\bibfield  {journal} {\bibinfo  {journal} {Phys. Rev. Lett.}\ }\textbf
  {\bibinfo {volume} {95}},\ \bibinfo {pages} {157203} (\bibinfo {year}
  {2005})}\BibitemShut {NoStop}%
\bibitem [{\citenamefont {Hwang}\ \emph {et~al.}(2019)\citenamefont {Hwang},
  \citenamefont {Wei}, \citenamefont {Huelga},\ and\ \citenamefont
  {Plenio}}]{Hwang2019}%
  \BibitemOpen
  \bibfield  {author} {\bibinfo {author} {\bibfnamefont {M.~J.}\ \bibnamefont
  {Hwang}}, \bibinfo {author} {\bibfnamefont {B.~B.}\ \bibnamefont {Wei}},
  \bibinfo {author} {\bibfnamefont {S.}~\bibnamefont {Huelga}}, \ and\ \bibinfo
  {author} {\bibfnamefont {M.~B.}\ \bibnamefont {Plenio}},\ }\href
  {https://arxiv.org/abs/1904.09937} {\bibfield  {journal} {\bibinfo  {journal}
  {arXiv preprint arXiv:1904.09937}\ } (\bibinfo {year} {2019})}\BibitemShut
  {NoStop}%
\bibitem [{\citenamefont {Wang}\ \emph {et~al.}(2015)\citenamefont {Wang},
  \citenamefont {Liu}, \citenamefont {Imri{\v{s}}ka}, \citenamefont {Ma},\ and\
  \citenamefont {Troyer}}]{Wang2015}%
  \BibitemOpen
  \bibfield  {author} {\bibinfo {author} {\bibfnamefont {L.}~\bibnamefont
  {Wang}}, \bibinfo {author} {\bibfnamefont {Y.-H.}\ \bibnamefont {Liu}},
  \bibinfo {author} {\bibfnamefont {J.}~\bibnamefont {Imri{\v{s}}ka}}, \bibinfo
  {author} {\bibfnamefont {P.~N.}\ \bibnamefont {Ma}}, \ and\ \bibinfo {author}
  {\bibfnamefont {M.}~\bibnamefont {Troyer}},\ }\href {\doibase
  10.1103/PhysRevX.5.031007} {\bibfield  {journal} {\bibinfo  {journal}
  {Physical Review X}\ }\textbf {\bibinfo {volume} {5}},\ \bibinfo {pages}
  {031007} (\bibinfo {year} {2015})}\BibitemShut {NoStop}%
\bibitem [{\citenamefont {Cai}\ \emph {et~al.}(2019)\citenamefont {Cai},
  \citenamefont {Hu}, \citenamefont {Ingersent}, \citenamefont {Paschen},\ and\
  \citenamefont {Si}}]{Cai2019}%
  \BibitemOpen
  \bibfield  {author} {\bibinfo {author} {\bibfnamefont {A.}~\bibnamefont
  {Cai}}, \bibinfo {author} {\bibfnamefont {H.}~\bibnamefont {Hu}}, \bibinfo
  {author} {\bibfnamefont {K.}~\bibnamefont {Ingersent}}, \bibinfo {author}
  {\bibfnamefont {S.}~\bibnamefont {Paschen}}, \ and\ \bibinfo {author}
  {\bibfnamefont {Q.}~\bibnamefont {Si}},\ }\href
  {https://arxiv.org/abs/1904.11471} {\bibfield  {journal} {\bibinfo  {journal}
  {arXiv preprint arXiv:1904.11471}\ } (\bibinfo {year} {2019})}\BibitemShut
  {NoStop}%
\bibitem [{\citenamefont {Gu}\ and\ \citenamefont {Yu}(2014)}]{Gu2014}%
  \BibitemOpen
  \bibfield  {author} {\bibinfo {author} {\bibfnamefont {S.~J.}\ \bibnamefont
  {Gu}}\ and\ \bibinfo {author} {\bibfnamefont {W.~C.}\ \bibnamefont {Yu}},\
  }\href {https://doi.org/10.1209/0295-5075/108/20002} {\bibfield  {journal}
  {\bibinfo  {journal} {Europhys. Lett. 108, 20002}\ } (\bibinfo {year}
  {2014})}\BibitemShut {NoStop}%
\bibitem [{\citenamefont {You}\ and\ \citenamefont {He}(2015)}]{You2015}%
  \BibitemOpen
  \bibfield  {author} {\bibinfo {author} {\bibfnamefont {W.~L.}\ \bibnamefont
  {You}}\ and\ \bibinfo {author} {\bibfnamefont {L.}~\bibnamefont {He}},\
  }\href {https://doi.org/10.1088/0953-8984/27/20/205601} {\bibfield  {journal}
  {\bibinfo  {journal} {J. Phys.: Condens. Matter 27, 205601}\ } (\bibinfo
  {year} {2015})}\BibitemShut {NoStop}%
\bibitem [{\citenamefont {Luo}\ \emph {et~al.}(2019)\citenamefont {Luo},
  \citenamefont {Zhao},\ and\ \citenamefont {Wang}}]{Luo2019}%
  \BibitemOpen
  \bibfield  {author} {\bibinfo {author} {\bibfnamefont {Q.}~\bibnamefont
  {Luo}}, \bibinfo {author} {\bibfnamefont {J.}~\bibnamefont {Zhao}}, \ and\
  \bibinfo {author} {\bibfnamefont {X.}~\bibnamefont {Wang}},\ }\href
  {https://arxiv.org/abs/1906.06553} {\bibfield  {journal} {\bibinfo  {journal}
  {arXiv preprint arXiv:1906.06553}\ } (\bibinfo {year} {2019})}\BibitemShut
  {NoStop}%
\end{thebibliography}%

\end{document}